\documentclass{emulateapj}
\usepackage{apjfonts}
\usepackage{lscape}
\input{epsf}

\shorttitle{Late-K and M dwarfs in the LAMOST commissioning data.}
\shortauthors{Zhong et al.}

\begin{document}
\title{AUTOMATED IDENTIFICATION OF 2,612 LATE-K AND M DWARFS
   	IN THE LAMOST COMMISSIONING DATA USING CLASSIFICATION TEMPLATE FITS}

\author{Jing Zhong\altaffilmark{1,2}, S\'ebastien
  L\'epine\altaffilmark{3,4,5}, Jinliang Hou\altaffilmark{1}, Shiyin
  Shen\altaffilmark{1}, Haibo Yuan\altaffilmark{6},Zhiying
  Huo\altaffilmark{6},Huihua Zhang\altaffilmark{6}, Maosheng
  Xiang\altaffilmark{6},Huawei Zhang\altaffilmark{6},Xiaowei
  Liu\altaffilmark{6,7}}

\altaffiltext{1}{Key Laboratory for Research in Galaxies and
  Cosmology, Shanghai Astronomical Observatory, Chinese Academy of
  Sciences, 80 Nandan Road, Shanghai 200030, China; jzhong@shao.ac.cn}

\altaffiltext{2}{University of Chinese Academy of Sciences, No. 19A,
  Yuquan Road, Beijing 100049, China}

\altaffiltext{3}{Department of Physics \& Astronomy, Georgia State
  University, 25 Park Place, Atlanta, GA 30303, USA;
  slepine@chara.gsu.edu}

\altaffiltext{4}{Department of Astrophysics, Division of Physical
  Sciences, American Museum of Natural History, Central Park West at
  79th Street, New York, NY 10024, USA}

\altaffiltext{5}{City University of New York, The Graduate Center, 365
  Fifth Avenue, New York, NY 10016, USA}

\altaffiltext{6}{Department of Astronomy, Peking University, Beijing
  100871, China}

\altaffiltext{7}{Kavli Institute for Astronomy and Astrophysics,
  Peking University, Beijing 100871, China }

\begin{abstract}
We develop a template-fit method to automatically identify and
classify late-type K and M dwarfs in spectra from the Large Sky Area
Multi-Object Fiber Spectroscopic Telescope (LAMOST). A search of the
commissioning data, acquired in 2009-2010, yields the identification
of 2,612 late-K and M dwarfs. The template fit method also provides
spectral classification to half a subtype, classifies the stars along
the dwarf-subdwarf (dM/sdM/esdM/usdM) metallicity sequence, and
provides improved metallicity/gravity information on a finer
scale. The automated search and classification is performed using a
set of cool star templates assembled from the Sloan Digital Sky Survey
spectroscopic database. We show that the stars can be efficiently
classified despite shortcomings in the LAMOST commissioning data which
include bright sky lines in the red. In particular we find that the
absolute and relative strengths of the critical TiO and CaH molecular
bands around 7000\AA are cleanly measured, which
provides accurate spectral typing from late-K to mid-M, and makes it
possible to estimate metallicity classes in a way that is more efficient and
reliable than with the use of spectral indices or spectral-index based
parameters such as $\zeta_{TiO/CaH}$. Most of the cool dwarfs observed
by LAMOST are found to be metal-rich dwarfs (dM). However, we identify
52 metal-poor M subdwarfs (sdM), 5 very metal-poor extreme subdwarfs
(esdM) and 1 probable ultra metal-poor subdwarf (usdM). We use a
calibration of spectral type to absolute magnitude and estimate
spectroscopic distances for all the stars; we also recover proper
motions from the SUPERBLINK and PPMXL catalogs. Our analysis of the
estimated transverse motions suggests a mean velocity and standard
deviation for the UVW components of velocity to be: <U>=-9.8 km/s,
$\sigma_{U}$=35.6 km/s; <V>=-22.8 km/s,$\sigma_{V}$=30.6 km/s;
<W>=-7.9 km/s, $\sigma_{W}$=28.4 km/s. The resulting values are
general agreement with previous reported results, which yields
confidence in our spectral classification and spectroscopic distance
estimates, and illustrates the potential for using LAMOST spectra of K
and M dwarfs for investigating the chemo-kinematics of the local
Galactic disk and halo.
\end{abstract}


\section{Introduction}

M dwarfs are the dominant class of stars in the Galaxy, and comprise
$\approx$70\% of all hydrogen-burning objects
\citep{Reid.2002,Covey.2008,Bochanski.2010}. Metal-rich M dwarfs and
their metal-poor counterparts, the M subdwarfs, hold great potential
for mapping out the baryonic mass and uncovering the dynamical
structure of the Galactic disk and halo. Their spectral energy
distribution is also very sensitive to metallicity variations, which
can potentially be used to map out the chemical evolution of the
Galaxy \citep{Bochanski.2013}. However, because M dwarfs are
relatively dim stars, with absolute visual magnitudes 8$<$M$_V<$15
\citep{Lepine.2005a}, and because current large-scale spectroscopic
surveys are efficient only for stars with magnitudes $V<20$, the
effective distance range over which M dwarfs can be studied in very
large numbers only extends to $\sim1-2$ Kpc.

Large multi-object spectroscopic surveys such as the Sloan Extension
for Galactic Understanding and Evolution \citep[SEGUE,][]{Yanny.2009}
have made considerable progress in uncovering the large-scale
structure of the Galaxy. This was achieved by targeting specific
tracers of distant stellar populations such as main-sequence turn-off
stars \citep{newberg.2002}. Chemical abundance and kinematic surveys
were also performed by targeting main sequence stars of intermediate
mass such as F and G stars \citep{Lee.2011}. A spectroscopic catalog
of M dwarfs identified in the Sloan Digital Sky Survey
\citep[SDSS,][]{York.2000}, and comprising more than 70,000 stars, was
presented by \citet{West.2011}, which showed that the metallicity of
early type M dwarfs decreases as a function of vertical distance from
Galactic plane.

As in SDSS/SEGUE, M dwarfs also make excellent targets for large
multi-object spectroscopic surveys such as LAMOST
\citep{Cui.2012}. This is because M dwarfs are very abundant in every
direction on the sky, and especially at low Galactic latitudes where
extra-galactic and Galactic halo surveys have low target
densities. They can thus provide abundant targets for dedicated
stellar surveys, and can also be used as filler targets for other
programs.

Since large numbers of M dwarf spectra are therefore expected to be
produced in the LAMOST survey, it is important to develop automated
and reliable tools to identify them and provide basic physical
properties such as temperature and metallicity. We are currently
developing a spectroscopic analysis pipeline to classify M dwarfs and
the more metal-poor M subdwarfs. Our goal is to provide fine-scale
spectral subtypes and molecular band ratios in order to place the
stars on a temperature-metallicity grid. This paper outlines the basic
method we have adopted, which is based on the use of high
signal-to-noise classification templates assembled from previous
spectroscopic surveys. To test the analysis pipeline, we have searched
for M dwarfs/subdwarfs in the LAMOST commissioning data, using M
dwarf/subdwarf templates assembled from the SDSS spectroscopic
survey. We present here a catalog of 2,612 M dwarfs and M subdwarfs
identified using our automated, template-fit method, and perform
simple distance and kinematics analysis which highlights the potential
of LAMOST for the study of low-mass stars and Galactic dynamics.

\section{The LAMOST commissioning data}

The Large sky Area Multi-Object fiber Spectroscopic Telescope
(LAMOST), also named the Guo Shou Jing Telescope, is a quasi-meridian
reflecting Schmidt telescope which provides a field of view of up to
20 square degrees, over which up to 4,000 optical fibers can be
automatically positioned, feeding light to 16 multi-object
spectrographs. Because of the large aperture optics design, the
telescope can produce 4,000 spectra in a single exposure, and is
designed to reach a limiting magnitude of about r=19 in 90 minutes
exposures, for spectra with a resolution R=1800
\citep{zhao.2012}. With an average expected rate of 6 exposures per
night, LAMOST will generate millions of spectra from stars and
galaxies every year, making it a powerful survey telescope.

A commissioning survey was implemented to test the capabilities of the
LAMOST telescope and verify its ability to target and observe 4000
stars at a time. To test the operation in survey mode, 6 fields have
been selected at moderately low Galactic latitudes, which provide an
abundance of bright targets. The fields are distributed near the
Galactic anti-center, with 119$^{\circ}$ $\lesssim gl  \lesssim $
226$^{\circ}$  and -25$^{\circ}$ $\lesssim  gb \lesssim $
+37$^{\circ}$. Each of the field was observed 4-16 times in the course
of the commissioning survey, each time targeting a different set of
4,000 stars in each field.

In this testing phase, however, only about 3500 of the 4000 fibers
were available for science targets. Some fibers were offline or could
not be moved at the time for a number of reasons, while other fibers
were trained on patches of background sky to measure local sky
brightness, and evaluate the contamination from sky lines from both
natural and artificial sources. The 3500 fibers available for science
targets were used on different objects in every exposure, as much as
possible, although a few stars ended up being observed twice.

In the end, a total of 165,219 spectra from science targets were
collected from the 48 field exposures. However, only a small fraction
of the spectra were found to have sufficient signal-to-noise ratio to
identify them as stars. Other spectra were dominated by sky
lines. This was found to be due to inefficiencies in the initial
LAMOST fiber positioning algorithm, which resulted in significant

Because of various instrumental and calibration problems, the
absolute wavelength calibration in the commissioning data cannot
be determined to better than a few angstrom for any star.
This issue has been improved in the pilot survey and regular
survey, e.g., \citealt{zhao.2012,Cui.2012}, and the typical error
of the  radial velocity is about 5 km s $^{-1}$ (Luo et al. in prep) .

Despite a few shortcomings in the test observations, the LAMOST
commissioning spectra still produced useful carry out some scientific
research, such as the discovery of 17 new planetary nebulae in the
edge of the M31\citep{Yuan.2010}, 14 new quasars near M31
\citep{Huo.2010}, 8 new quasars in the extragalactic field
\citep{Wu.2010a,Wu.2010b}, and 9 candidate metal-poor stars with
[Fe/H] $\le$ -1.0 \citep{Li.2010}. This suggest that the LAMOST
commissioning data maybe has a sufficient quality for us to perform
the M dwarfs/subdwarfs identification and classification.

Table~\ref{table1} summarizes information on the six fields that were
observed, including the coordinates of the field centers, the number
of times each field was surveyed, and the mean signal-to-noise ratio
for the spectra obtained in the field. The mean signal-to-noise ratio
is calculated by averaging the photon counts from all the science
fibers, which include those where only sky lines were detected
(i.e. fibers targeting stars but for which the alignment was
incorrect, resulting in only sky being detected). The quoted S/N
values therefore underestimate the true potential of LAMOST.

%
%
\begin{deluxetable}{crrcr}
\tabletypesize{\scriptsize}
\tablecolumns{5}
\tablewidth{0pt}
\tablecaption{Summary of the commissioning observation\label{table1}}
\tablehead{\colhead{FIELD} & \colhead{RA(2000)\tablenotemark{1}} & \colhead{DEC(2000)}
& \colhead{N\_obs\tablenotemark{2}} & \colhead{S/N\tablenotemark{3}} }
\startdata
I & 91.028979 & 23.264102 & 16 & 14.39 \\
II & 124.75370 & 56.254898 & 11 & 9.08 \\
III & 18.136639 & 45.335527 & 7 & 11.42 \\
IV & 42.884911 & 35.057388 & 5 & 12.77 \\
V & 11.162321 & 40.679614 & 5 & 9.09 \\
VI & 124.40772 & 0.477676 & 4 & 13.94
\enddata
\tablenotetext{1}{Center point of celestial coordinates for the field.}
\tablenotetext{2}{Observed frequency of field.}
\tablenotetext{3}{Mean signal-to-noise ratio of field.}
\end{deluxetable}

\section{M dwarf search and classification}

\subsection{Spectral Templates}

\begin{figure*}[t]
\vspace{0cm}
\hspace{0cm}
\includegraphics[angle=90,scale=.6]{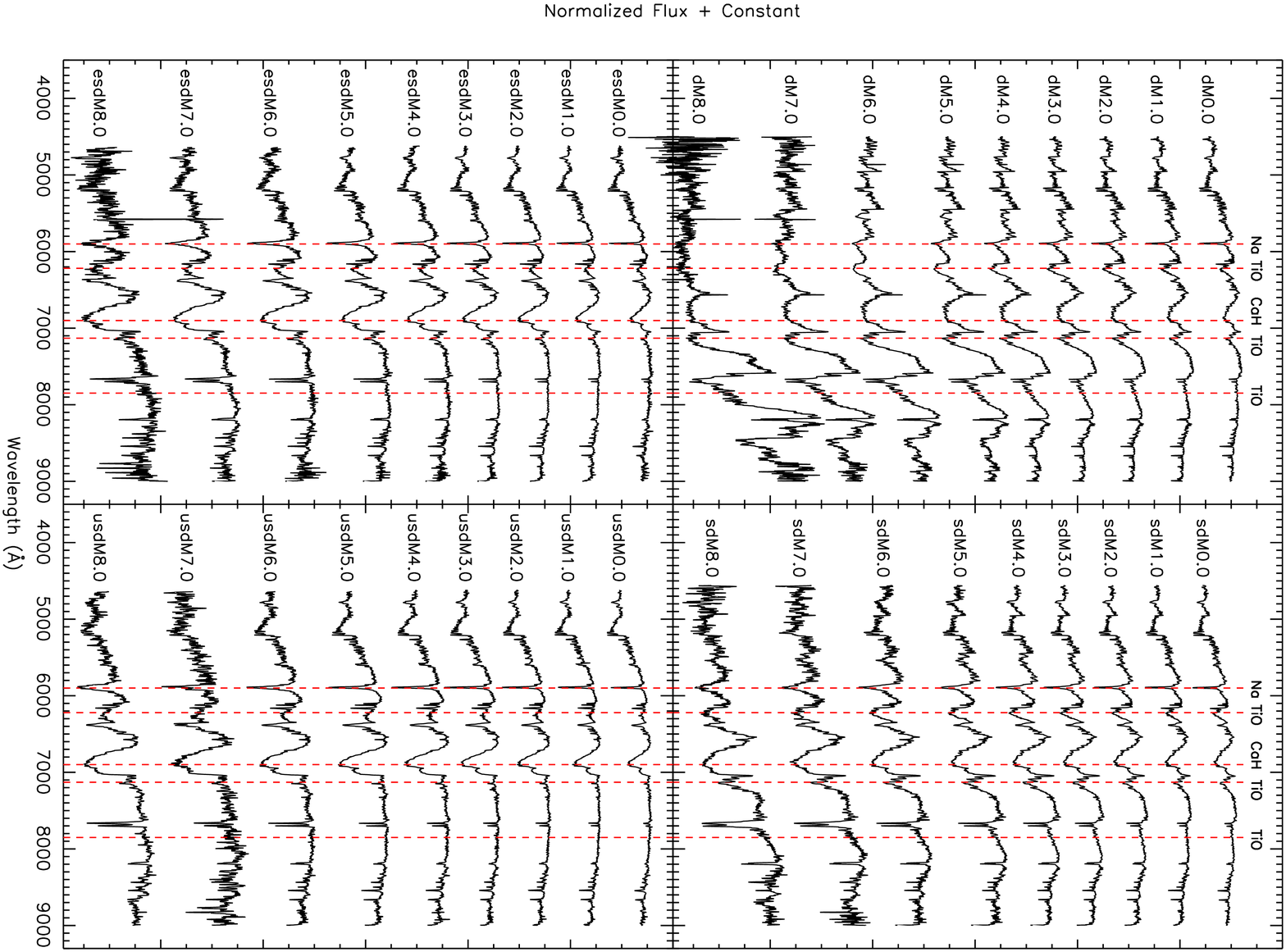}
\caption{Sequence of synthetic classification templates assembled from
  SDSS spectra. Each of the four sets above shows a temperature
  sequence, with M0.0 stars being the warmest and M8.0 stars the
  coolest in the sequence. The four classes (dwarfs:dM;
  subdwarfs:sdM; extreme subdwarfs:esdM; and ultrasubdwars:usdM)
  allegedly represent a metallicity sequence, with the dwarfs (dM)
  being the most metal rich and the ultrasubdwarfs (usdM) the most
  metal-poor. Late-type templates are noisier due to the smaller
  number and higher signal-to-noise of the SDSS spectra that were
  co-added to generate the templates. \label{templates_T}}
\end{figure*}

\begin{figure*}[t]
\vspace{0cm}
\hspace{0cm}
\includegraphics[angle=90,scale=.6]{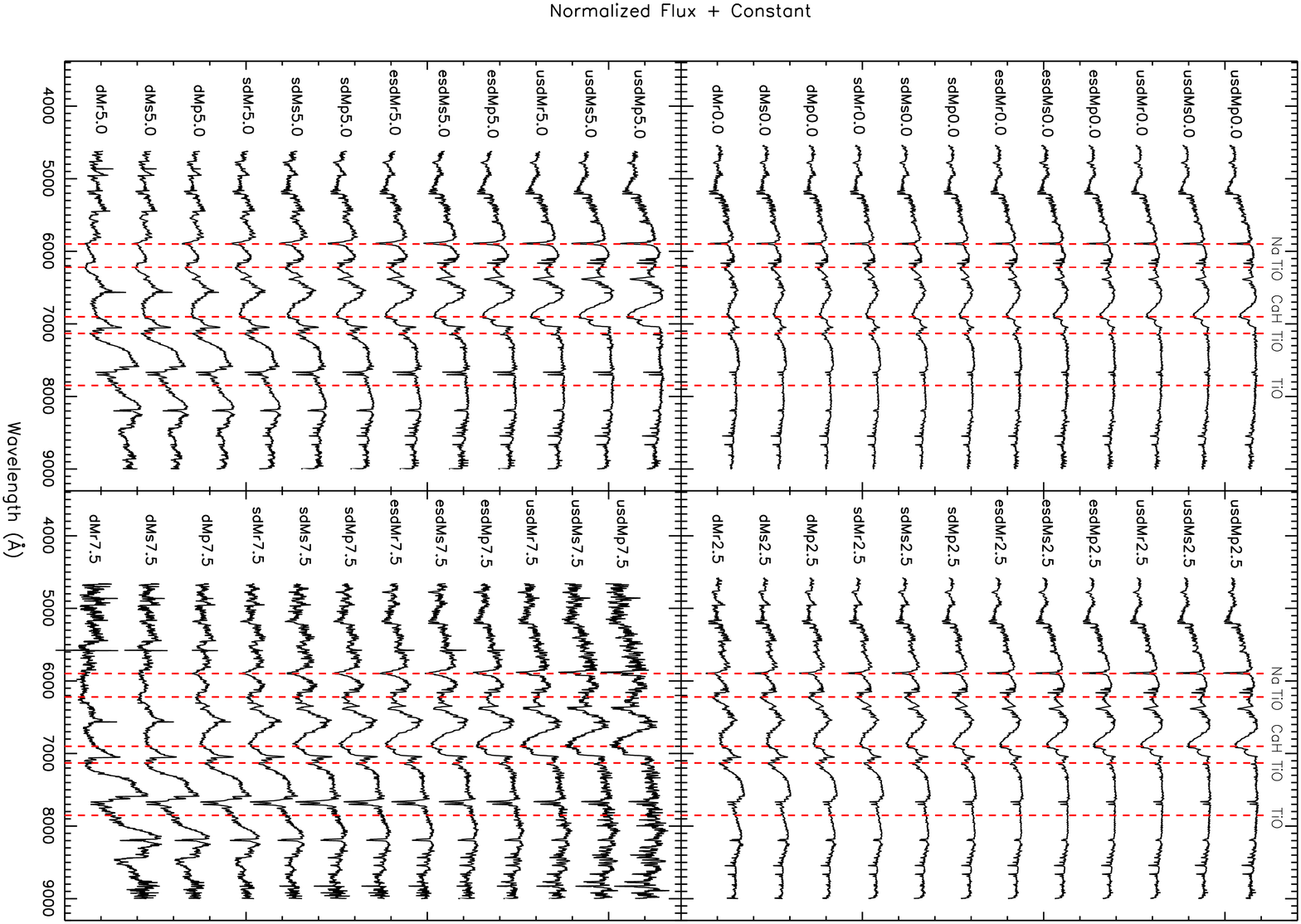}
\caption{Subsets of classification templates for stars of similar
  subtypes but different metallicities. We define twelves different
  metallicity subclasses, all shown here. These range from dMr to
  usdMp, from the most metal-rich to the most metal poor. Four of the
  temperature subtypes (0.0, 2.5, 5.0, 7.5) are selected to illustrate
  the difference of metallicity classes. The late subtype templates
  (7.5) are noisier because of the smaller number and lower S/N of the
  SDSS spectra that were combined to generate
  them.\label{templates_M}}
\end{figure*}

Formal spectral classification of M dwarfs is, in principle, based on a
direct comparison with classification standards that have been
assembled over the years
\citep{Mould.1976,Bessell.1982,kirkpatrick.1991,kirkpatrick.1995,Bochanski.2007}. Formal
sets of primary and secondary classification standards for the M
subdwarf (i.e. metal-poor) sequences have also been proposed in
\citet{Lepine.2007}. Visual classification is however not practical
for large spectroscopic surveys where thousands of spectra are
collected every night.

Automated classification methods have been developed over the years
based on measures of the strength of TiO, CaH, and VO bands, which
dominate the optical spectra of M dwarfs. Spectral indices were
introduced by \citet{Reid.1995} and \citet{Lepine.2003a,Lepine.2007}
and relationships between values of theses indices and spectral
subtypes have been calibrated, such that subtypes can in principle be
estimated from spectral index measurements alone; a process that can
be automated. A drawback of the technique is that artifacts in the
spectra (sky lines, cosmic rays, calibration errors) can result in
significant errors in the spectral subtype estimates. In addition,
differences in spectral resolution and spectrophotometric calibration
from different telescopes/instruments can result in systematic errors
in the measurement of the spectral indices, which require careful
recalibrations \citep{Lepine.2013}.

To help in the visual verification of automated spectral
classification results, a specialized software was developed for the
Sloan Digital Sky Survey, called "The Hammer"\citep{covey.2007}. The
Hammer performs spectral classification using a set of spectral
indices which assign a tentative subtype. Spectra can then be
displayed on a graphical interface and allow a user to identify errors
and correct the spectral typing from visual inspection. Using this
method, \citet{West.2011} visually inspected over 70,000 M dwarfs from
the SDSS DR7. The method still requires considerable investment of
human resources however, but was applied with some success in a search
for late-type stellar spectra in LAMOST datasets \citep{Yi.2014}.

As an alternative to these classification methods, we are proposing to
develop a procedure for automated classification which provides
reliable spectral subtypes without prior measurement of spectral
indices, and without the need for visually inspecting every
spectrum. Our method rather relies on direct fits to spectral
classification templates assembled by combining spectra of M dwarfs
and M subdwarfs identified in the Sloan Digital Sky Survey. Our
general philosophy is to identify and classify M dwarfs using all the
available spectroscopic information within the most relevant
wavelength range for these stars, by performing a fit to identify the
template which best matches the data.

Our templates were assembled from SDSS spectra of relatively bright
(r<18) late-K and M dwarfs, most drawn from the subset of M dwarfs
released in the SDSS DR7, and presented in \citet{West.2011}. In the
catalog presented by \citet{West.2011}, the M dwarfs were classified
using the Hammer code, with subtypes listed to the nearest
integer; instead of using these subtypes, we re-classified all the
stars based on the spectral index method described in
\citet{Lepine.2003a,Lepine.2007}, which is based on measurements of
the three spectral band indices TiO5, CaH2, and CaH3, which measure
the strengths of the TiO and CaH molecular bands near 7000\AA. In a
first pass, the spectral indices were measured,
the stars classified, and spectra were co-added to produce an initial
set of dM/sdM/esdM/usdM classification templates. Atomic lines from
the Ca II triplet ($\lambda$= 8498, 8542, 8662\AA) were
measured on the templates to determine any systematic radial velocity
shift, and the templates were shifted back to the stellar rest
frame. In a second pass, each SDSS spectrum was cross-correlated with
its matching template, and the radial velocity shift for each star was
measured (generally to a precision of +/- 10 km/s). Spectral indices
were then re-measured for each star shifted back in the local rest
frame, and each star was re-classified. A new set of classification
templates was then generated. This procedure was repeated for a third
pass. After the third pass, the final co-adds were selected as formal
classification templates; the final number of SDSS spectra used to
define each classification template is shown in Table~\ref{table2}.

\begin{deluxetable}{lrrrr}
\tabletypesize{\scriptsize}
\tablewidth{200pt}
\tablecolumns{5}
\tablecaption{Number of SDSS spectra combined to make the
  classification templates \label{table2}}
\tablehead{
\colhead{} &
\colhead{d...} &
\colhead{sd...} &
\colhead{esd...} &
\colhead{usd...}
}
\startdata
...K7.0 & 1,755 & 12 & 11 & 13 \\
...K7.5 & 1,870 & 12 & 16 & 14 \\
...M0.0 & 1,651 & 17 & 20 & 29 \\
...M0.5 &   785 & 14 & 18 & 10 \\
...M1.0 &   688 & 22 & 20 & 17 \\
...M1.5 &   843 & 17 & 25 & 17 \\
...M2.0 &   601 & 18 & 20 & 11 \\
...M2.5 &   711 & 21 & 16 &  8 \\
...M3.0 &   299 & 41 & 24 &  8 \\
...M3.5 &   813 & 27 & 20 &  8 \\
...M4.0 &   813 & 21 &  6 &  8 \\
...M4.5 &   643 &  6 &  7 &  2 \\
...M5.0 &   643 &  4 &  2 &  5 \\
...M5.5 &   778 &  4 &  3 &  2 \\
...M6.0 &   478 &  2 &  2 &  2 \\
...M6.5 &    26 &  1 &  5 &  1 \\
...M7.0 &    31 &  1 &  2 &  1 \\
...M7.5 &    29 &  2 &  2 &  1 \\
...M8.0 &    20 &  2 &  2 & \nodata\\
...M8.5 &    16 &  2 & \nodata & \nodata
\enddata	
\end{deluxetable}

A subset of the final templates is presented in Figures
~\ref{templates_T} and ~\ref{templates_M}. The
resulting templates span the spectral subtypes K7.0-M8.5, and cover
every half-subtype in those ranges. These initial templates are
arranged in a two-dimensional grid, with one axis measuring the
general strength of the molecular bands and the other axis measuring
the ratio of the TiO and CaH band strengths. The first axis (spectral
subtype) is presumed to be mainly correlated with a star's effective
temperature. The second axis, on the other hand, is presumed to be
mainly correlated with a star's metallicity, although this need not be
strictly the case; for instance, the ratio of TiO to CaH can also be a
function of surface gravity. In any case, we will assume that our
classification system essentially represents a
"temperature-metallicity" grid. With over 18 elements of resolution in
"temperature", and 4 elements of resolution in "metallicity". (We note
that the term "metallicity" is only used here to suggest that the
variations in TiO to CaH band strength ratios are most likely due to
variations in a star's chemical composition. It is clear that these
specific ratios may not be strictly correlated with the abundance of
iron, so our use of the word "metallicity" here is not meant to
suggest that values of [Fe/H] can be inferred from those grids.) The
limited resolution along the "metallicity" axis is due in part to the
difficulty in measuring accurate metallicity values in individual
stars using the TiO5 and CaH2+CaH3 spectral indices alone, combined
with the limited signal-to-noise ratio in the individual stars. Our
co-added templates however provide significantly higher
signal-to-noise, which make it possible to define a finer "metallicity"
scale.

To increase the number of metallicity grid points from 4 to 12, we
used the following procedure. For any one of the four original
metallicity classes (i.e., dM), each can be expanded to three
subclasses: a relatively more metal-rich subclass (i.e., dMr), a
standard subclass (i.e. dMs), and a relatively more metal-poor
subclass (i.e., dMp). Assuming that the metallicity grid is a linear
system \citep[see more details in ][]{Mann.2012}, the extra synthetic
templates can be created by extrapolating between two neighboring
classes. For example, we can synthesize dMp templates by considering
them to be one third of the way from the dM standard class to the sdM
standard class:
\begin{equation}
[dMp] =\frac{[sdMs]-[dMs]}{3}+[dMs]=\frac{2}{3}[dMs]+\frac{1}{3}[sdMs]
\label{eqdmp}
\end{equation}
where dMs and sdMs are the original template spectra of the dwarfs and
subdwarfs. We can thus interpolate between dMs and sdMs to create dMp
and sdMr subclass templates, between sdMs and esdMs to create sdMp and
esdMr subclass templates, and between esdMs and usdMs to create esdMp
and usdMr subclass templates. To create the dMr and usdMp templates,
we can extrapolate them from the original templates assuming a linear
progression at the edges. To make the dMr template, we can assume that
the dMs template is between dMp and dMr, which means
$dMs=([dMp]+[dMr])/2$, and using the Eq. \ref{eqdmp}, the dMr template
can be derived:
\begin{equation}
[dMr] =2[dMs]-[dMp]=\frac{4}{3}[dMs]-\frac{1}{3}[sdMs]
\end{equation}
And the usdMp subclass template can be similarly synthesized by the
same extrapolation from the other edge.

\begin{figure}[t]
\vspace{0cm}
\hspace{.5cm}
\includegraphics[angle=90,scale=0.6]{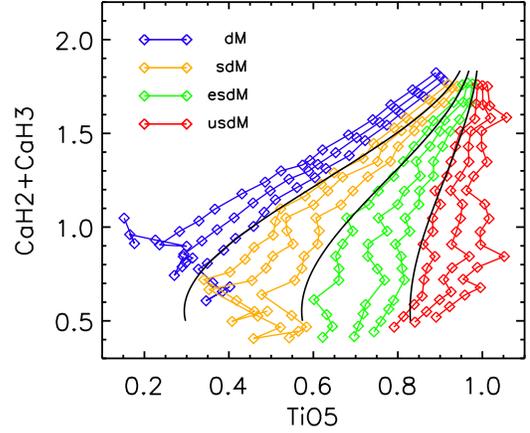}
\caption{Distribution of the TiO5 and CaH2+CaH3 spectral indices in
  the classification templates. The four colors represent the four
  original metallicity classes: dwarfs (blue), subdwarfs (yellow),
  extreme subdwarfs (green), and ultrasubdwarfs (red). Our templates
  further subdivide each metallicity class into three subclasses:
  relatively more metal-rich subclass(Mr), standard subclass(Ms), and
  relatively more metal-poor subclass(Mp), from left to right on the
  plot, respectively. Indices for the three subclasses are
  shown. Segments connect the index values of templates from the
  different spectral subtypes within each of the metallicity
  subclasses. The boundaries of the four metallicity subclasses, as
  proposed by \citet{Lepine.2013} are shown as black lines. Although
  our templates were originally assembled by the \citet{Lepine.2007}
  calibration, which makes a small offset between a few early-type
  synthetic templates and the boundary lines, the differences were not
  significant. Overall, our synthetic templates define a finer
  metallicity grid, while still following the original subclass
  separation.\label{idx_tmp}}
\end{figure}

To verify the validity of our synthetic method, the distribution of
TiO5 and CaH2+CaH3 spectral indices from all the synthetic templates
is shown in Figure~\ref{idx_tmp}. Four different colors represent the
four original metallicity classes, which are dwarfs (blue), subdwarfs
(yellow), extreme subdwarfs (green), and ultra-subdwarfs (red), from
left to right on the plot, respectively. For each metallicity class
color, the middle line is the spectral index distribution of the
corresponding original template sequence, while the leftmost line
corresponds to the relatively more "metal-rich" subclass sequence (r),
and the rightmost line corresponds to the relatively more "metal-pool"
subclass sequence (p). The three black lines are the boundaries
proposed by \citet{Lepine.2013} to separate the four original
metallicity classes. This shows that our synthetic templates not only
define a finer "metallicity" grid but also still follow the original
classification criteria within each metallicity subclass. However, for
later subtypes (dM7.0-dM8.5, sdM7.0-sdM8.5), the relative rarity of
SDSS spectra having appropriate "temperature" and "metallicity" values
results in the templates being dominated by the spectra of just a few
stars, which can skew the mean "metallicity" and "temperature" of the
subsample, due to fluctuations from small number statistics. On the
other hand, the low S/N ratio of late-type template spectra also leads
to larger errors in the measurement of spectral indices and limits
their value as a subtype/metallicity test
(Figure~\ref{idx_tmp}). Therefore, metallicity class estimates at the
latest subtypes based on these templates should be used with
caution.

We emphasize here again that we use the terms "temperature" and
"metallicity" in our classification system as an educated suggestion
for what we believe underlies the two dimensional classification
grid. These are not meant to suggest that each grid point represents a
unique combination of the physical parameters Teff and [Fe/H]. What
this means is that this proposed classification system should be used
for classification purposes, and not as a means to determine physical
parameters for a star. The determination of physical parameters from
this classification would first require a validation and calibration
of the grid. This is in fact an ongoing area of research in the field
of cool stars which has seen some progress in recent year, see
e.g.,\cite{2012ApJ...748...93R, 2013AJ....145...52M, 2014AJ....147..160M, 2015ApJ...800...85N}.
 It should be understood in the following section that the
term "temperature" is used in the sense of "spectral subtype", the
first axis in our classification grid, and that the term "metallicity"
is used to described the second axis in our classification system,
which really describes molecular band ratios.

\begin{figure}[t]
\vspace{0cm}
\hspace{0.8cm}
\includegraphics[angle=90,scale=0.48]{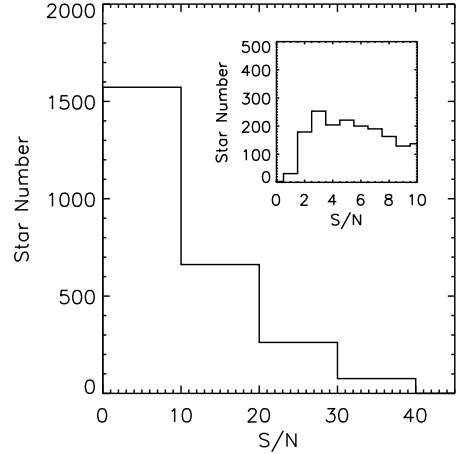}
\caption{Distribution of estimated mean S/N ratio in the LAMOST
  spectra for 2,612 M dwarfs/subdwarfs. The sub-panel display the mean
  S/N distribution in the range $0<$S/N$<10$.\label{his_sn}}
\end{figure}

\begin{figure*}[t]
\begin{center}
\vspace{0.0cm}
\hspace{0cm}
\includegraphics[angle=90,scale=0.39]{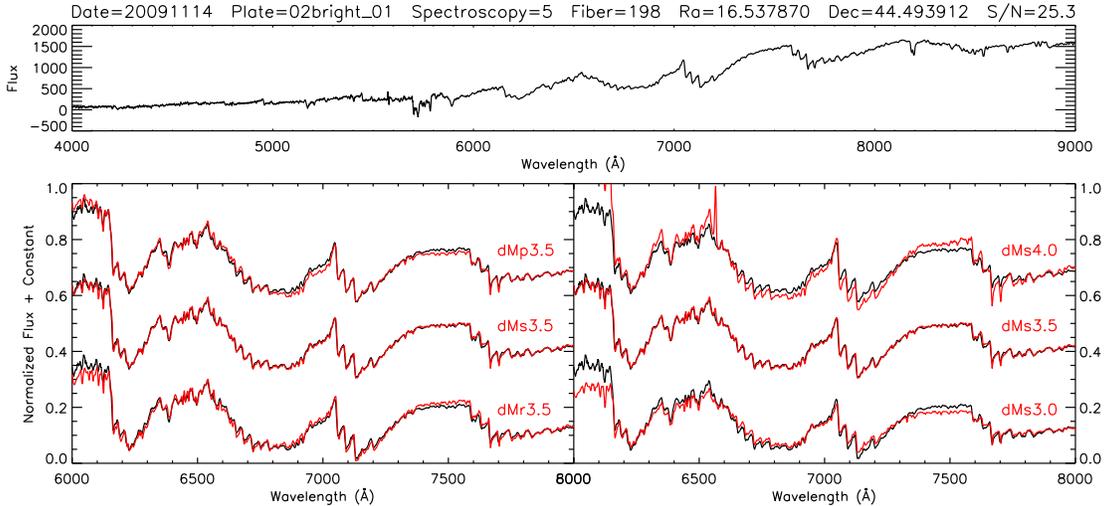}
\caption{Quality of the fit for a representative high S/N spectrum in
  the LAMOST commissioning data, identified to be an M3.5 dwarf by our
  template-fitting method. {\it Top panel}: the full spectrum, which
  is a composite of the LAMOST blue channel and red channel
  spectra. {\it Bottom panels}: the smoothed LAMOST spectrum (black)
  overlaid on classification templates selected to demonstrate the
  quality of the fit (red). {\it Bottom left}: Three adjacent
  templates of the same subtype but different metallicity
  subclasses. The middle one is the best fitting template; the one
  above is the relatively more metal-poor template, and the one below
  is the relatively more metal-rich template, as labeled. The
  difference between top/bottom templates and the LAMOST spectrum is
  clear in the 6000\AA--6200\AA, 6900\AA--7100\AA (CaH and TiO band)
  and 7400\AA--7600\AA (VO band) ranges. {\it Bottom right}: Three
  adjacent templates of the same metallicity subclass but different
  spectral subtypes. The middle one is the best fitting template,
  while those above/below represent templates of higher/lower
  temperatures in the sequence. The most significant difference is in
  the 6000\AA -- 6200\AA range.\label{dm3.5}}
\end{center}
\end{figure*}

\begin{figure*}[t]
\begin{center}
\vspace{0cm}
\hspace{0.cm}
\includegraphics[angle=90,scale=0.39]{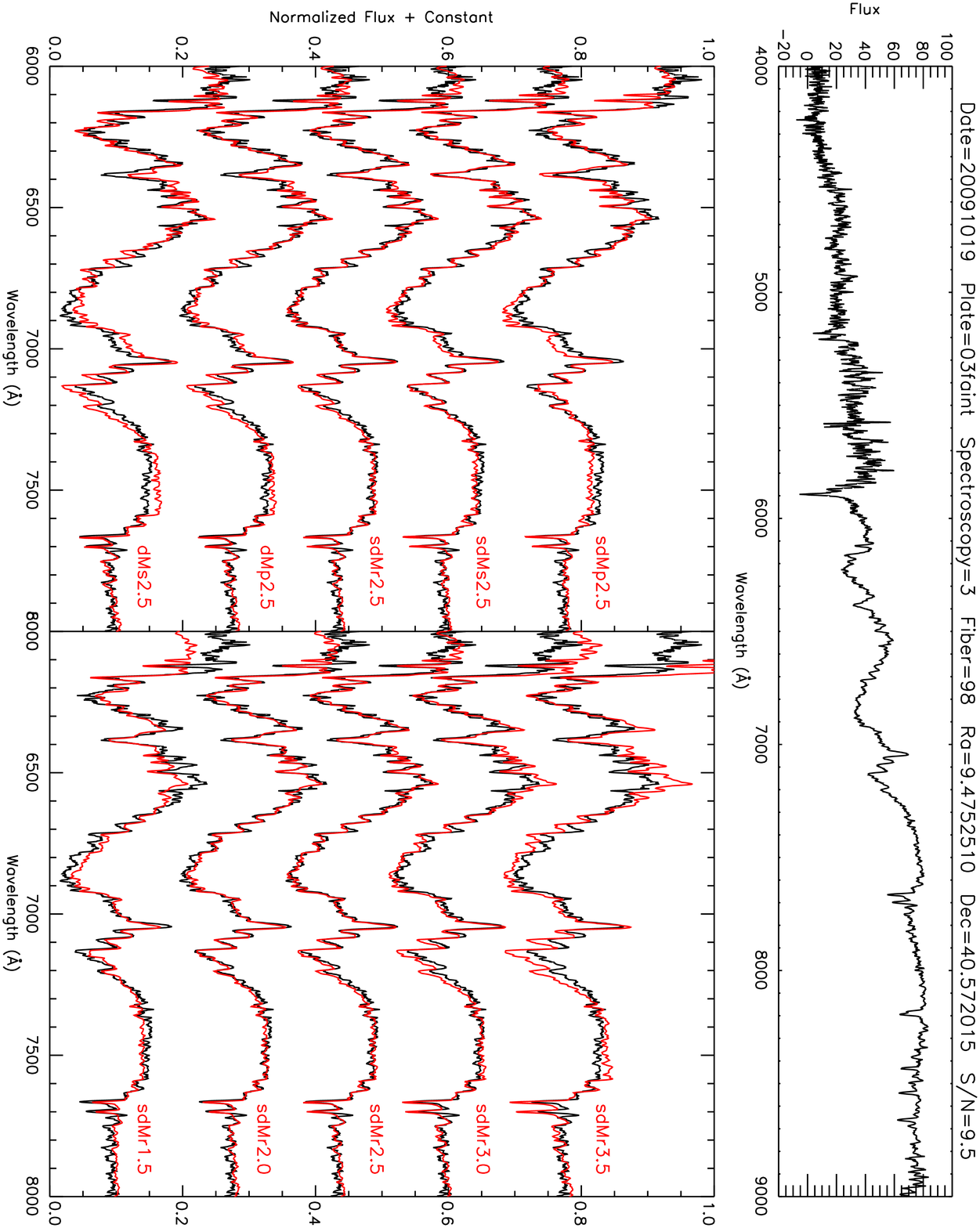}
\caption{Quality of the fit for a representative LAMOST spectrum of
  intermediate S/N. The star happens to be classified as sdMr2.5 by
  our pipeline. Panels are the same as in Figure~\ref{dm3.5}, except that
  additional templates extending $\pm$1 metallicity subclasses and
  $\pm$1.0 spectral subtypes from the best fit are also shown for
  comparison.\label{sdm2.5}}
\end{center}
\end{figure*}

\begin{figure*}[t]
\begin{center}
\vspace{0cm}
\hspace{0.cm}
\includegraphics[angle=90,scale=0.39]{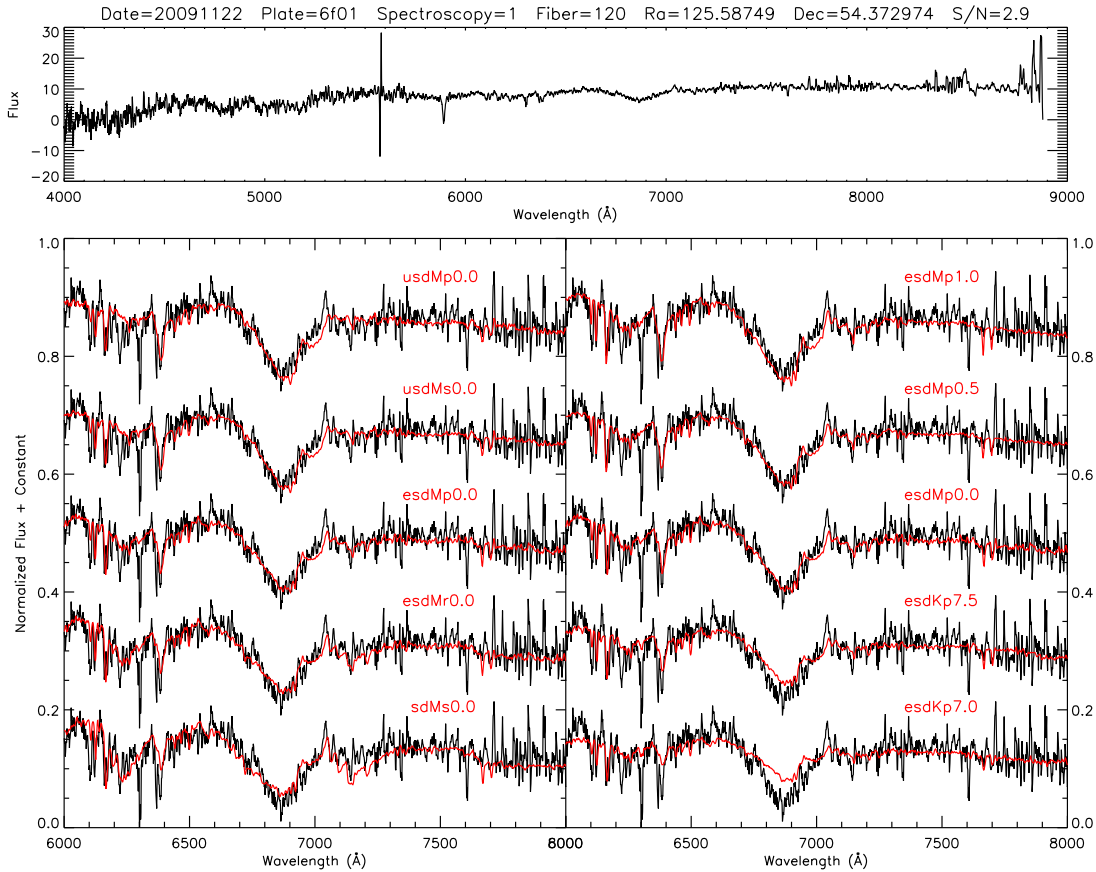}
\caption{Quality of the fit for a typical LAMOST spectrum of low
  S/N. The star happens to be classified as an extreme subdwarf
  (esdMp0.0) by our pipeline. Because of the serious skylines
  contamination and the weakness of the molecular bands, it is hard
  for our template-fit method to classified the spectrum with the same
  accuracy as the higher S/N spectra shown in Figure~\ref{dm3.5} and
  Figure~\ref{sdm2.5}. In any case, close examination of neighboring
  templates (above and below the best fit template shown in the
  middle) reveals difference that are significant enough near the
  7,000\AA region to be confident of a classification accuracy to
  within $\pm$1.0 spectral subtypes and within $\pm$2 metallicity
  subclasses. \label{esdm0.0}}
\end{center}
\end{figure*}

\begin{figure*}[t]
\begin{center}
\vspace{0cm}
\hspace{0.cm}
\includegraphics[angle=90,scale=0.55]{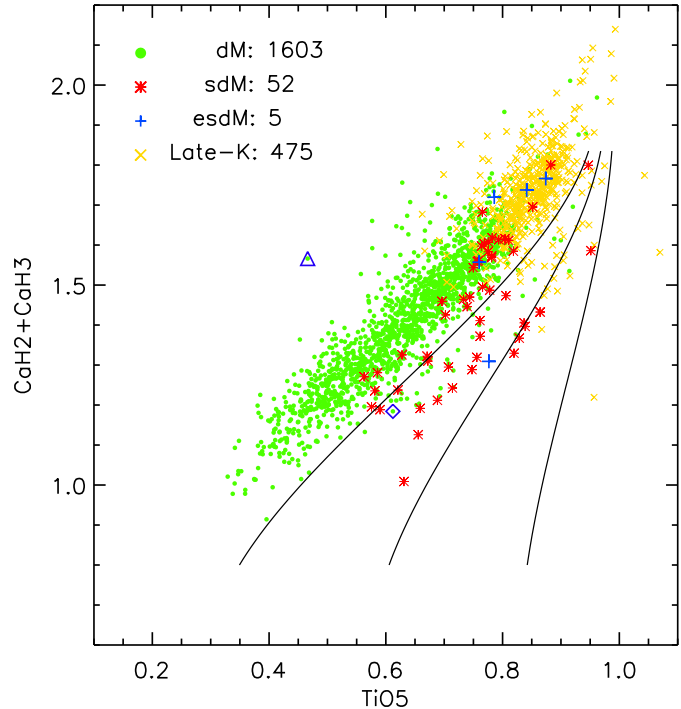}
\caption{
  Distribution of the spectral index measuring the bandhead of the
  TiO band redward of 7040\AA (TiO5) and the sum of the indices
  measuring the strength of the CaH bandhead blueward of 7040\AA
  (CaH2+CaH3). Different symbols are used for stars classified in the
  LAMOST classification pipeline as dwarfs (dM), subdwarfs (sdM),
  extreme subdwarfs (esdM), and late-K dwarfs. As expected, many of
  the subdwarfs and extreme subdwarfs fall to the right of the dM dwarfs
  sequence, due to the TiO/CaH band ratio being metallicity dependent.
  Some stars classified as M subdwarfs have indices more consistent
  with dM dwarfs; those stars tend to have lower S/N and could be
  misclassified, or suffer significant measurement errors on the
  spectral indices. The three solid black lines delimitate the loci of
  the four metallicity classes (dM, sdM, esdM, usdM, left to right)
  originally proposed by \citet{Lepine.2013}. Two outliers, classified
  as dM dwarfs but falling outside the normal distribution of dM stars
  in this diagram (open blue triangle and diamond symbols), are
  investigated in Figure~\ref{exp}. \label{idx_all}}
\end{center}
\end{figure*}

\begin{figure*}[t]
\begin{center}
\hspace{0cm}
\vspace{0cm}
\includegraphics[angle=90,scale=0.45]{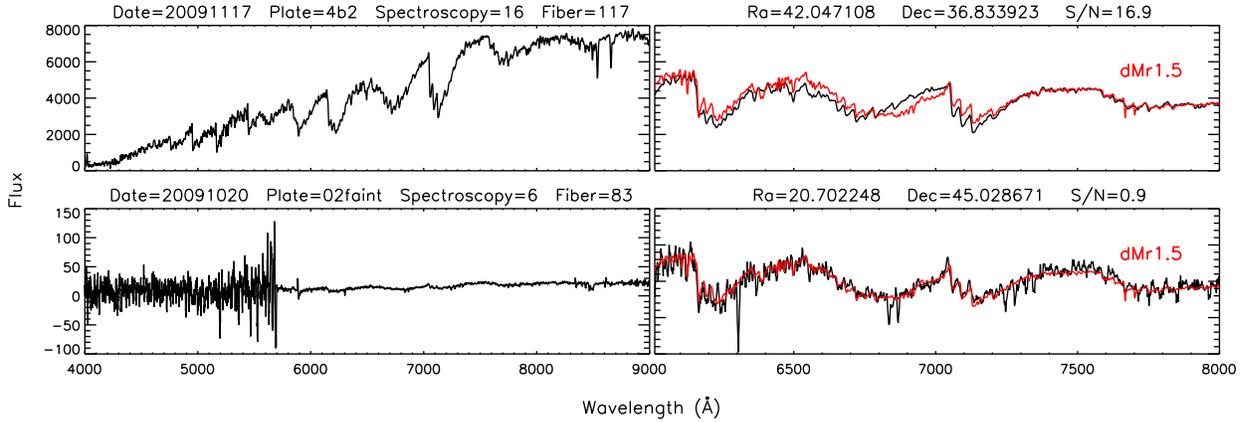}
\caption{Spectra of the two outlier stars marked in
  Figure~\ref{idx_all}. $Top$: spectrum of the outlier identified as
  an open triangle symbol. The weaker molecular bands of CaH,
  consistent with the location in the TiO/CaH diagram of
  Figure~\ref{idx_all}, indicate that it is most probably an M giant
  instead of an M dwarf. $Bottom$: spectrum of the outlier represented
  by an open diamond symbol. Because of the low S/N ratio, the
  skylines occurring in the CaH molecular bands result in significant
  errors in the spectral index measurements. The template-fit
  classification appears to be quite good however.\label{exp}}
\end{center}
\end{figure*}

\begin{figure*}[t]
\begin{center}
\vspace{0cm}
\hspace{0.cm}
\includegraphics[angle=90,scale=0.55]{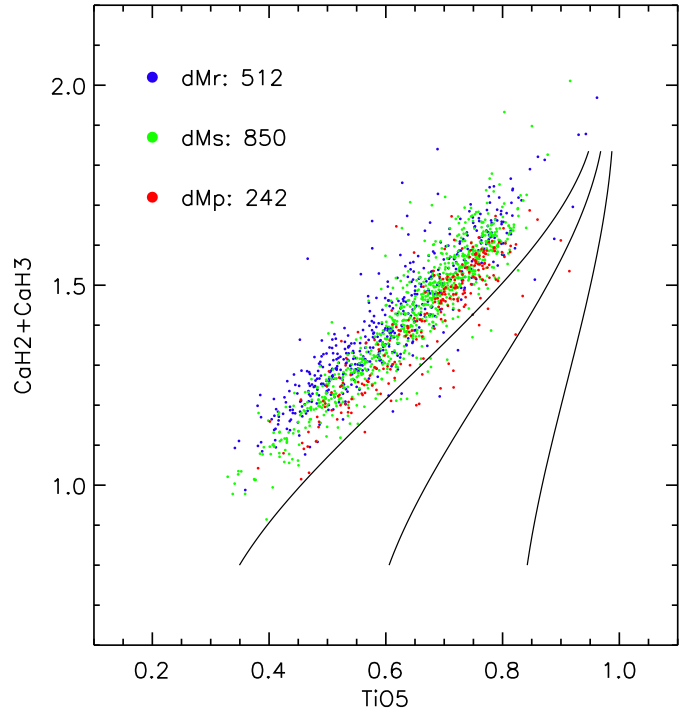}
\caption{Spectral index distribution of dM dwarfs for the three
  different metallicity subclasses defined in this paper. The
  different loci of the dMp, dMs, and dMr subclasses clearly show
  that the metallicity subclasses defined in our spectral fitting
  method also consistently separate out the stars in the spectral
  index grid, according the expected trends for metallicity
  variations.\label{idx_dm}}
\end{center}
\end{figure*}

\begin{figure}[t]
\vspace{0cm}
\hspace{1cm}
\includegraphics[scale=0.8]{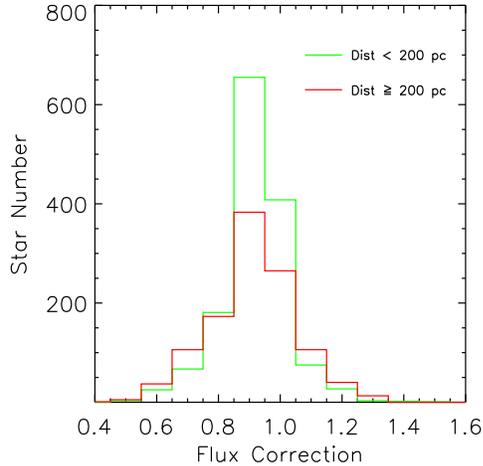}
\caption{Distribution of the flux correction factor $fc$, which
  measure whether the spectral energy distribution of a LAMOST M dwarf
  looks redder ($fc<1$) or bluer ($fc>1$) than the adopted SDSS-based
  classification templates. The distribution suggests that the flux
  calibration is generally good for most of the LAMOST commissioning
  spectra, albeit redder than expected on average. To investigate the
  possibility of interstellar reddening, we divide the stars in two
  groups based on their spectroscopic distance. The closer stars tend
  to have relatively fewer outliers but still look marginally redder
  than expected, on average, which points to differences in the LAMOST
  and SDSS flux calibrations.
\label{fc}}
\end{figure}

\begin{figure*}[t]
\vspace{0cm}
\hspace{1cm}
\begin{center}
\includegraphics[scale=0.65]{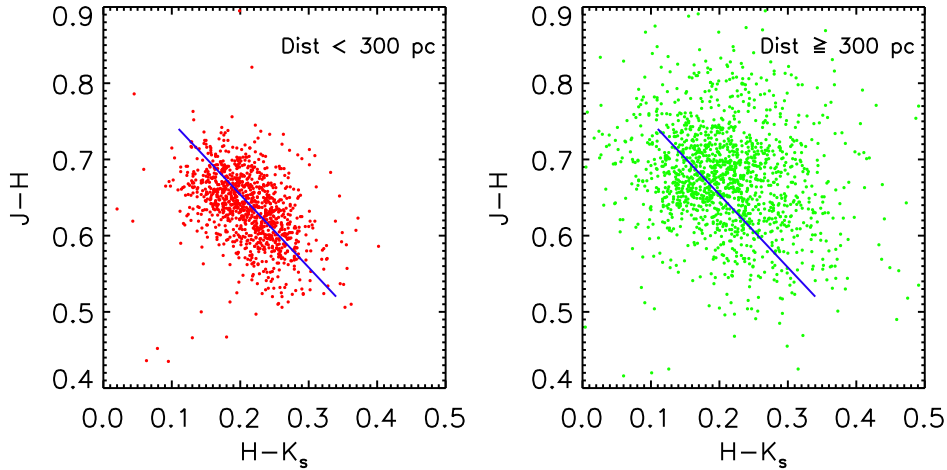}
\caption{Distribution of 2MASS J-H and H-K$_s$ colors for the stars
  identified as late-K and M dwarfs/subdwarfs in the LAMOST
  commissioning data. $Left$: the 1,080 nearest stars
  ($d<300pc$); their relative concentration is consistent with
  unreddened main sequence stars as shows in the infrared color-color
  diagram of \citet{Lepine.2005b}. This suggest that there is no
  significant interstellar reddening or red giants
  contamination. $Right$: the 1,497 more distant stars
  ($d>300pc$). The larger scatter is in part due to stars have fainter
  magnitudes, and thus larger uncertainty in their infrared colors. The
  distribution, however, also shows redder colors on average. This
  systematic offset suggests that some stars may suffer some form of
  interstellar reddening, which would be consistent with the larger
  distance range.\label{jh_hk}}
\end{center}
\end{figure*}

\begin{figure*}[t]
\begin{center}
\vspace{0cm}
\hspace{0.3cm}
\includegraphics[angle=0,scale=0.55]{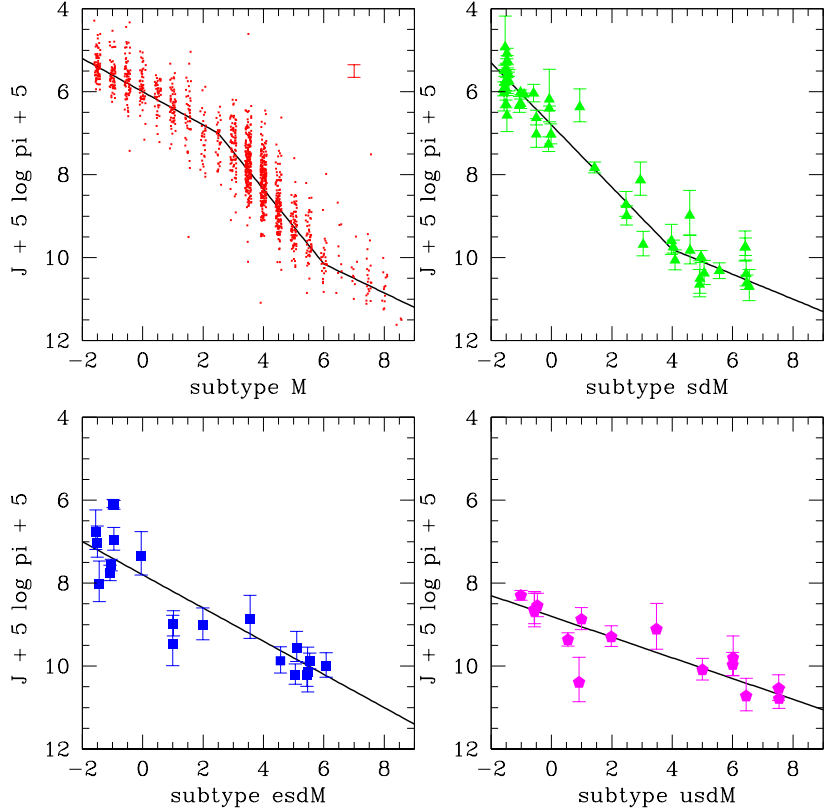}
\caption{Distribution of absolute magnitudes in J band for nearby M
  stars with both spectral subtypes and parallax distances. A random
  values of $\pm$ 0.1 subtype was added, to spread out the data for
  more clarity. Stars from different metallicity classes (dM, sdM,
  esdM, usdM) are shown in different panels. The black lines show the
  adopted relationships for each metallicity subclass.\label{mag_spty}}
\end{center}
\end{figure*}

\begin{figure}[t]
\vspace{0cm}
\hspace{0.3cm}
\includegraphics[angle=90,scale=0.5]{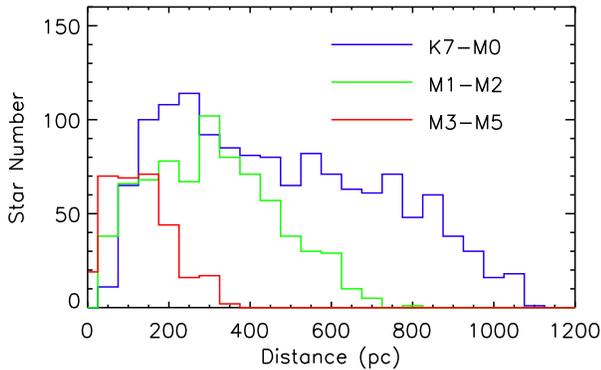}
\caption{Distribution of spectroscopic distances for stars in three
  spectral subtype ranges (K7-M0, M1-M2, and M3-M5). The early-type
  dwarfs are dominated in our catalog and are sampled over a larger
  spectroscopic distances than other late-type dwarfs. This bias is
  entirely due to magnitude selection effects, which are expected
  because the early-type M dwarfs are intrinsically brighter than the
  late-type M dwarfs.\label{dis_his}}
\end{figure}

\begin{figure}[t]
\begin{center}
\includegraphics[scale=0.5]{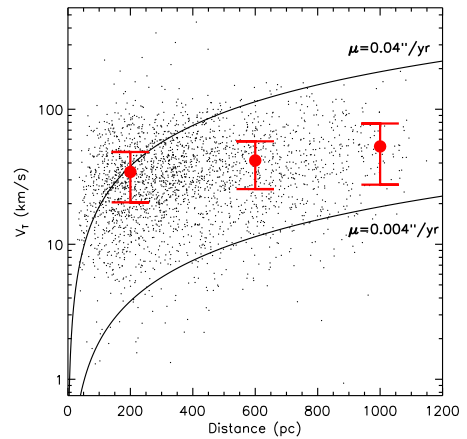}
\caption{Distribution of transverse velocities as a function of the
  spectroscopic distance. Mean velocities are calculated for stars in
  three distance bins (0-400 pc, 400-800 pc, 800-1200 pc); the red
  dots and error bars represent the median transverse velocity and the
  corresponding standard deviation for each bin. The mean transverse
  is largely consistent with the local kinematics of disk
  stars. However, it shows a weak growing trend with spectroscopic
  distance, which can be simply explained by the uncertainty in the
  spectroscopic distance estimates.\label{vt_dist}}
\end{center}
\end{figure}

\begin{figure*}[t]
\begin{center}
\includegraphics[angle=0,scale=0.65]{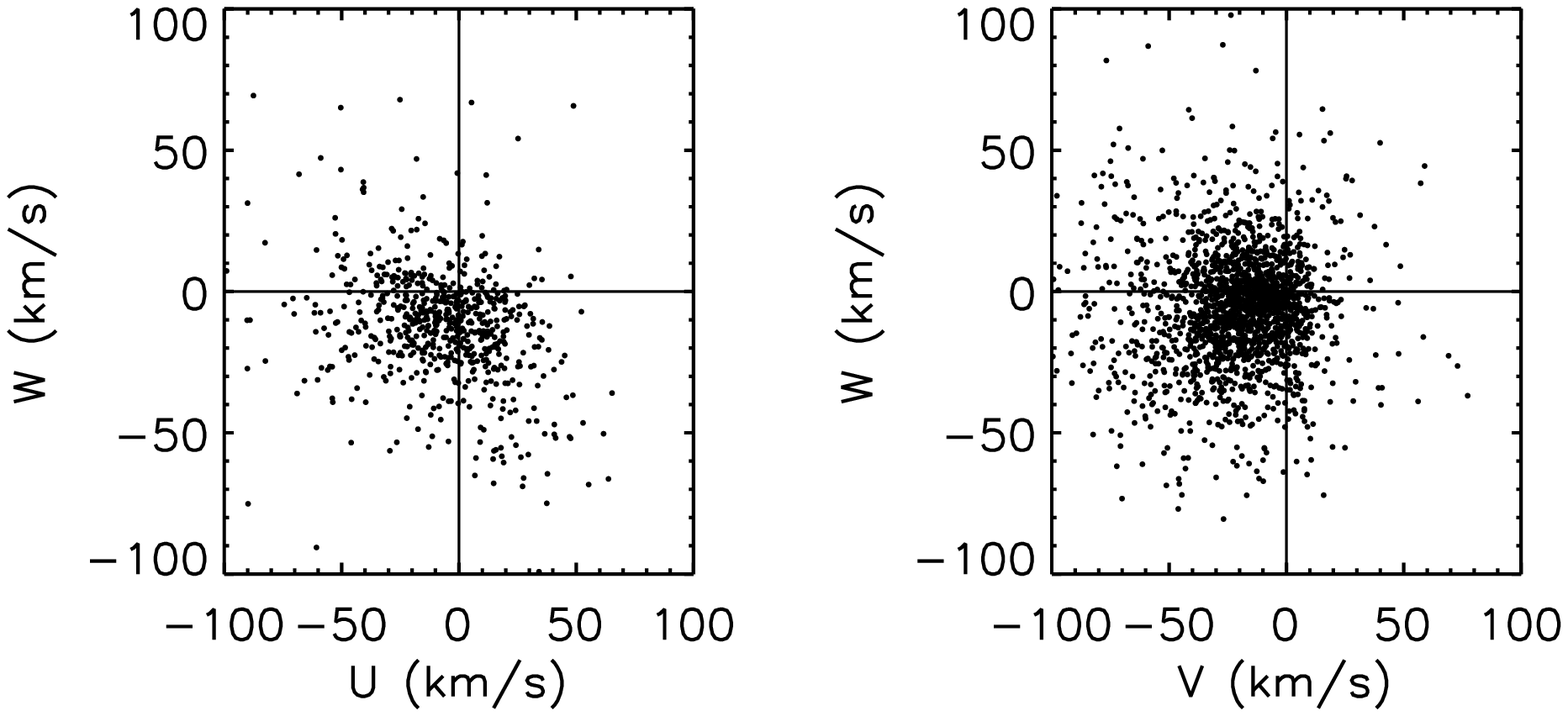}
\figcaption{Distribution of UVW components of motion for our
spectroscopically classified M dwarfs in LAMOST, based on
spectroscopic distances and proper motion measurements using the
method described in \S5.5. Each M dwarf contributes only two
velocity components, which means it is only shown in one of the
panels. The (V,W) projection shows the expected asymmetric drift of
local disk stars.\label{uvw}}
\end{center}
\end{figure*}

\subsection{Method for template fitting}

We perform our spectral classification by calculating the chi-square
value between any one LAMOST spectrum and each of the template
spectra. The spectra are first reduced using the LAMOST 2D pipeline
(\citealt{Luo.2004}), which includes bias subtraction, flatfield
correction, skyline subtraction, wavelength calibration, and flux
calibration (\citealt{zhao.2012}). For best results, we have
determined that the spectral fitting should avoid the overlap region
between the blue-channel and red-channel spectra, which is between
5700\AA and 5900\AA (\citealt{Wu.2011}), because of noise and flux
calibration issues. The spectral fit should also avoid the wavelength
range between 8000{\AA} -- 9000\AA where serious skyline contamination
occurs. The spectral wavelength range used for the classification fit
is thus limited to 6000\AA--8000\AA. Bad pixels, flagged by the LAMOST
reduction pipeline, are also excluded from the fit. The chi-square for
the $j$th template spectrum is defined as:
\begin{equation}
\chi_{j}^2=\sum_{i=1}^{N}\left(\frac{L_{i}-T_{ij}} {\sigma_{i}}\right)^{2}
\end{equation}
where $L_{i}$ is the LAMOST spectral flux for each of the i=1,N
pixels, $T_{ij}$ is the flux of the $j$th template spectrum for every
corresponding pixel $i$, and $\sigma_{i}$ is the error for the LAMOST
spectral flux at each pixel $i$. N is equal to 1250 because there are
1250 pixels between 6000\AA and 8000\AA.

In order to correct for radial velocity shifts, and reduce the
possible effects of faulty wavelength calibration (See \S 2 above),
we have to makesure that the stars are classified in their local rest frame as much
as possible. We therefore shift each LAMOST spectrum towards the blue
or red by an integer number of pixels, and recalculate the chi-square
value against each template spectrum. For each LAMOST spectrum, we allow
for a possible maximum shift of $\pm$8 pixels, which corresponds to
maximal radial velocity shifts of approximately $\pm$550 km/s. The
factor combination of pixel shifting and $j$th template spectrum which
gives the minimum chi-square value is selected as the best fit.

Stars are rejected when their best fit is a match to a non-M dwarf
template. For each of the 12 metallicity subclasses, there are two
earlier template spectra (called non-M dwarf templates) corresponding
to K5.0 and K5.5 dwarf/subdwarf subtypes; these templates have
extremely weak molecular features of titanium oxide (TiO), vanadium
oxide (VO) and calcium hydride (CaH). For stellar types earlier than
K7.0, these molecular bands are indeed so weak as to be generally
undiscernible, except in spectra with very high S/N, and the spectra
essentially all look relatively flat in the 6000\AA -- 8000\AA
range (\citealt{Reid.1995}). As a result, any star with a subtype
earlier than K7 (including all F and G stars) will have its best match
with any one of the K5.0 or K5.5 templates. Any star with a best match
to those templates is therefore identified as a non-M dwarf, and
rejected.

Since the quantum efficiency for CCDs in LAMOST is increasing over the
wavelength range from 6000\AA to 8000\AA, the measurement errors for
each pixel have a systematically decreasing trend over this wavelength
range. So when we use the LAMOST spectrum's chi-square fit to the
template, the red part close to 8000\AA has more weight than the blue
part closer to 6000\AA. On the other hand, most of the prominent
molecular absorption features are in this bluer part. This means that
some non-M dwarf stars which happen to have similar spectra on the red
side around 8000\AA might be classified as M dwarfs, even if there is
no good fit on the blue side. This is more common for low S/N spectra,
because their blue edge is dominated by the instrumental noise, and
their red edge is as featureless as most of the templates, and thus
they appear more similar to the early-M dwarf templates than to the
late-K ones.

To solve this problem, we use an additional criterion to identify and
reject non-M dwarfs with noisy spectra that might otherwise be
classified as M dwarfs from the chi-square fit. In addition to the
chi-square value, we also calculate the integrated square of the
difference between the LAMOST spectrum and the best classification
template. If the integrated square difference is greater than 130
pixel units, then the star is considered a non-conclusive
classification and is excluded from the final M dwarf
list. Specifically, this means that we allow for a maximum difference
between a LAMOST spectrum and a classification template of 10\% for
each pixel {\em on average}; for the normalized spectra, the
corresponding difference is 0.1 units. Because there are about 1250
pixels used in the fit, the threshold value in the fitting area is
about 130 pixel units. Our experience shows that this procedure
eliminates the vast majority of noisy, but clearly non-M dwarf stars
that may otherwise be identified as M dwarfs from the chi-square fit.

\subsection{Additional Spectrophotometric Correction}

We find that to obtain the best results with our template fit methods,
both the LAMOST spectra and the classification templates also need to
be flux-normalized before the fit, i.e. the spectra and templates need
to be rectified with respect to a pseudo-continuum level. This ensures
that the classification is primarily based on the absolute and
relative strength of the molecular bands, and not on the slope of the
spectral energy distribution, which can be affected by many factors
such as errors in instrumental flux calibration, aperture losses
combined with atmospheric differential refraction, or interstellar
reddening.

Considering that the molecular bands completely dominate the spectral
energy distribution within the fitting area, it is hard to objectively
define a continuum for rectification. We therefore use a more simple
flux recalibration that plays a similar role and performs a first
order correction of the slope.

For each spectrum, we divide the spectral fitting range in two parts,
and calculate the integrated flux within each of these two regions.
Then the straight line $y=ax+b$ that fits the two median flux points
is used to calibrate the flux and the slope of this spectrum. The
parameters $a$ and $b$ are given by: $a=(F_{\rm r}-F_{\rm
  b})/(\lambda_{\rm r}-\lambda_{\rm b})$, $b=F_{\rm b}-a\lambda_{\rm
  b} $,
where $F_{\rm r}$ is the median flux of the red section
(7000\AA-8000\AA), $F_{\rm b}$ is the median flux of the blue section
(6000\AA-7000\AA), and $\lambda_{\rm r}$, $\lambda_{\rm b}$ are the
median wavelengths within each of the two regions. The same flux
recalibration is performed on all the classification templates, and
the chi-square best fit is performed for the {\em normalized} spectrum
as it is fit on the {\em normalized} templates.

To evaluate the magnitude of this flux correction, we first define $R$
as the ratio of the spectral fluxes, $R=F_{\rm r}/F_{\rm b}$, where
$F_{\rm r}$, $F_{\rm b}$ are the median fluxes which have been defined
before. We then introduce $fc$ as the ratio of the red/blue flux ratio
R for the original LAMOST spectrum (i.e. before flux correction) and
the red/blue flux ratio R for the classification template which
provides the best fit to the data (with the fit performed after flux
recalibration). This factor of the flux correction $fc$ is thus
defined as:
\begin{equation}
  fc=\frac{R_{\rm LAMOST}} {R_{\rm SDSS}}
\end{equation}
where $R_{\rm LAMOST}$ is the spectral flux ratio of the LAMOST
spectrum and $R_{\rm SDSS}$ is the flux ratio of the classification
template. Using this flux correction factor, we can verify whether the
LAMOST flux calibration and SDSS flux calibration are similar
($fc=1$), indicating that the LAMOST spectrum has the same measured
spectral energy distribution as an SDSS spectrum of the same spectral
subtype. A value different from $fc=1$ would indicate that star has a
spectral energy distribution that is different than that of a
classification standard star of the same spectral type. The value of
$fc$ can thus be used, e.g., to diagnose reddening in a star, which
would produce a value of $fc<1$.

Again, the final template fit is performed {\em after} these
additional flux corrections. Stars which are found to match one of the
templates are then automatically assigned a metallicity class (dMr,
dMs, dMp, sdMr, sdMs, sdMp, esdMr, esdMs, esdMp, usdMr, usdMs, usdMp)
and spectral subtype (0.0 to 8.5) from the class and subtype of the
template which provided the best fit. We also include stars classified
with subtypes K7.0 and K7.5, and of all metallicity subclasses.

\section{Application and Verification of the Classification Method}

\subsection{Search results}

We performed spectral classification of all the spectra collected
during the LAMOST commissioning phase, using our template-fit
method. Of the 83,500 spectra which were passed through the M dwarf
classification pipeline, 2,612 were positively identified as M
dwarfs/subdwarfs by our classification code, 74,573 spectra (89.3\%)
were excluded as non-M dwarfs by the template fitting method, and
6,315 spectra (7.6\%) were rejected as too noisy for classification,
based on our integrated square-difference criterion.

Superficial examination of the spectra from the 2,612 M dwarf
candidates shows that all of them display TiO and CaH bands, which
confirms that they are all consistent with M dwarf/subdwarf
stars. Examination of some of the 80,888 rejected spectra, on the other
hand, reveals a variety of other sources, some looking like stars of
earlier spectral types such as AFGK stars and white dwarfs, others
showing only sky lines, or spectra that are too noisy to be
identified. In any case, none of the rejected spectra are found to
show clear evidence of TiO or CaH molecular bands, and are thus not M
stars.

\subsection{Sample spectra}

To verify the quality of the template fits and how effectively M
dwarfs/subdwarfs are classified with our pipeline, we performed close
visual inspection of a number of spectra covering a broad range of
spectral subtypes (as determined from our fit) and with a range of
integrated square differences to the classification template.

Since the classification accuracy is mainly influenced by the S/N
ratio for each spectrum, the S/N distribution for 2,612 M
dwarfs/subdwarfs is shown in Figure~\ref{his_sn}. Based on this
histogram, we can divide the M dwarfs/subdwarfs classified in our
pipeline into three subsets: the high S/N spectra ($S/N>15$), the
intermediate S/N spectra ($5<S/N<15$), and the low S/N spectra
($S/N<5$). To illustrate the fit quality of different S/N subsets, we
plot three example spectra and their fitting templates in
Figures~\ref{dm3.5}, \ref{sdm2.5}, and \ref{esdm0.0}.

Figure~\ref{dm3.5} displays the LAMOST high S/N spectrum of a star
classified as dM3.5 by our pipeline (black line). The spectrum is
compared to the best-fitting template (red line). The best-fitting
template (dM3.5) shows a near-perfect fit to the LAMOST spectrum, and
in particular is an excellent match to the main molecular absorption
bands (TiO, CaH) between 6000\AA and 8000\AA. To evaluate the quality
of the fit, we also compare this LAMOST spectrum with the four adjacent
templates in the spectral subtype / metallicity class grid, which are
overlaid on the LAMOST spectrum. There are subtle but clear differences
between each of these adjacent templates and the LAMOST spectrum,
which demonstrate that the class/subtype dM3.5 provides the best
possible classification for the star. This suggest that for
high-quality data, our classification method is indeed accurate at
least to the nearest half-subtype, and to the nearest metallicity
subclass.

Figure~\ref{sdm2.5} shows the case of a LAMOST spectrum of
intermediate S/N, which happens to be identified as an M subdwarf,
with a sdMr2.5 classification. Although the differences between
early-type dwarfs in the temperature-metallicity grid are not as clear
as for the dMs3.5 star in Figure~\ref{dm3.5}, there are still notable
differences which indicate that the fit is not as good for the
adjacent templates. With the differences between the LAMOST spectrum
and the templates increasing significantly for templates beyond those
displayed in Figure~\ref{sdm2.5}, we remain confident that the
classification also has an accuracy of at least +/-0.5 subtypes, and
is accurate to within the nearest metallicity subclass.

Figure~\ref{esdm0.0} shows the low S/N spectrum of a LAMOST star,
identified as an extreme subdwarf with a classification esdMp0.0
according to our code. The weakness of the molecular bands and the
lower S/N makes it harder for the star to be accurately classified,
and there is more uncertainty in the assigned spectral type and
metallicity class. In any case, there are significant differences
between the star and templates that are two steps away from the
assigned subtype/subclass. From this we are confident that the
classification is still accurate to within +/- 1.0 subtypes, and to
within $\pm$2 metallicity subclasses.

Our pipeline has classified 2,612 M dwarfs/subdwarfs in total. The
accuracy for low S/N spectra classification are no better than $\pm$1.0
subtype or $\pm$2 metallicity subclasses as defined in this paper,
however these low S/N spectra (S/N <5) are a minority and only
represent 25.8\% of the catalog (675 stars). The remaining 1,937 stars
have spectra with sufficient S/N ratio (>5) to be accurately
classified to within $\pm$0.5 subtype or $\pm$1 metallicity subclass.

\subsection{Measurement of spectral indices}

As an additional verification of our classification method, we have
calculated sets of spectral indices commonly used for M dwarf/subdwarf
classification. The TiO5, CaH2, and CaH3 spectral indices, as defined
in \citet{Reid.1995}, were calculated for all the LAMOST stars
identified as M dwarfs/subwarfs by our classification pipeline.

The distribution of TiO5 against CaH2+CaH3 was originally used to
formally separate metal-poor M subdwarfs from metal-rich M dwarfs
\citep{Gizis.1997,Lepine.2003a,Lepine.2007}. We plot this distribution
for all the LAMOST M dwarfs/subdwarfs in Figure~\ref{idx_all}, using
different symbols/colors for stars of different metallicity
classes. We note that the distribution is in good agreement with
Figure 3 of \citet{Lepine.2007}, and confirms that most of the LAMOST
stars are relatively metal-rich M dwarfs and late-K dwarfs. The LAMOST
stars classified as subdwarfs generally have higher TiO5 index value
for a given range of CaH2+CaH3 values, also in general agreement with
the spectral index classification system.

However, there are several outliers, e.g., stars classified as dM by
our pipeline but with a [TiO,CaH2+CaH3] index more consistent with an
sdM subtype. We carefully examine two of these to understand the
discrepancy. The two stars represent two different kinds of offset
along the dM sequence. The top panel of Figure~\ref{exp} displays the
spectrum of the star identified as an open triangle symbol in
Figure~\ref{idx_all}. The difference in the molecular absorption bands
between the star and any of the dM templates suggests that it is
probably an M giant instead of M dwarf; the weaker molecular bands of
CaH are consistent with the M giant spectra analyzed by
\citet{Mann.2012}. In this particular example, it is our classification
pipeline that is in error, and the TiO/CaH spectral index ratio more
correctly identifies the star as M giant. It is hard for our
automated classification pipeline to exclude these M giant stars with
the template-fit method, because the dwarf templates always provide
the best fit to the spectra and the molecular bands are similar. Our
best option would be to use M giant templates in addition to the M
dwarfs/subdwarfs. Future efforts will be devoted to adding M giant
templates to the pipeline.

The second outlier, represented by a diamond symbol in
Figure~\ref{idx_all}, is plotted in the bottom panels of
Figure~\ref{exp}. Here the source of the offset is found to be due to
instrumental factors. The low S/N ratio and intense skyline
contamination in this spectrum cause a large error in the measurement
of the spectral indices. In this case, it is the spectral index
measurements that are in error, and our template fit method provides a
more reliable estimate of the subtype and metallicity class.

More generally, we find that the spectral-fit method provides a better
estimate of the spectral subtype because it uses a much broader
wavelength range, whereas the TiO5, CaH2, and CaH3 spectral indices
are narrowly defined, and thus much more sensitive to instrumental
noise or spectral reduction artefacts. We find that large scatter of
the sdM stars in Figure~\ref{idx_all} is thus largely due to errors in
the measurement the spectral indices from low S/N spectra, not in
errors in our spectral classification pipeline. We conclude suggests
that our classification pipeline is more reliable for measuring the
spectral type and the metallicity than if we were using the spectral
indices alone.

To verify the consistency of our newly introduced metallicity
subclasses (r,s,p), we plot the spectral index distribution of dM
dwarfs for the three different subclasses in Figure~\ref{idx_dm}. The
slightly different locations (layering) of the dMp, dMs, and dMr stars
suggest that the metallicity subclasses defined in our spectral
fitting method also consistently separate out the stars in the
spectral index grid following the expected trends for metallicity
variations, with the more metal-rich stars (dMr) having lower TiO5
values for a given CaH2+CaH3, and the more metal-poor stars (dMp)
having higher TiO5 values.

\begin{deluxetable}{lrrrr}
\centering
\tabletypesize{\scriptsize}
\tablewidth{200pt}
\tablecolumns{5}
\tablecaption{ The absolute magnitude ($M_{J}$) for spectral subtypes (SpTy)
\label{table3}}
\tablehead{
\colhead{} &
\colhead{d...} &
\colhead{sd...} &
\colhead{esd...} &
\colhead{usd...}
}
\startdata
  ...K7.0   &    5.600  &   6.050  &   7.400  &   8.550 \\
  ...K7.5   &    5.800  &   6.425  &   7.600  &   8.675 \\
  ...M0.0   &    6.000  &   6.800  &   7.800  &   8.800 \\
  ...M0.5   &    6.200  &   7.175  &   8.000  &   8.925 \\
  ...M1.0   &    6.400  &   7.550  &   8.200  &   9.050 \\
  ...M1.5   &    6.600  &   7.925  &   8.400  &   9.175 \\
  ...M2.0   &    6.800  &   8.300  &   8.600  &   9.300 \\
  ...M2.5   &    7.000  &   8.675  &   8.800  &   9.425 \\
  ...M3.0   &    7.450  &   9.050  &   9.000  &   9.550 \\
  ...M3.5   &    7.900  &   9.425  &   9.200  &   9.675 \\
  ...M4.0   &    8.350  &   9.800  &   9.400  &   9.800 \\
  ...M4.5   &    8.800  &   9.950  &   9.600  &   9.925 \\
  ...M5.0   &    9.250  &  10.100  &   9.800  &  10.050 \\
  ...M5.5   &    9.700  &  10.250  &  10.000  &  10.175 \\
  ...M6.0   &   10.150  &  10.400  &  10.200  &  10.300 \\
  ...M6.5   &   10.325  &  10.550  &  10.400  &  10.425 \\
  ...M7.0   &   10.500  &  10.700  &  10.600  &  10.550 \\
  ...M7.5   &   10.675  &  10.850  &  10.800  &  10.675 \\
  ...M8.0   &   10.850  &  11.000  &  11.000  &  10.800 \\
  ...M8.5   &   11.025  &  11.150  &  11.200  &  10.925 \\
  ...M9.0   &   11.200  &  11.300  &  11.400  &  11.050 \\
  ...M9.5   &   11.375  &  11.450  &  11.600  &  11.175
\enddata	
\end{deluxetable}

\subsection{Flux correction}

As we defined in Section 3.3, the flux correction, $fc$, is mostly
useful to verify the quality of the LAMOST flux
calibration. Figure~\ref{fc} shows the distribution of $fc$ values for
all 2,612 stars classified as M dwarfs/subdwarfs. The distribution
suggests that the flux calibration is in general agreement with the
SDSS calibration for most of the LAMOST commissioning
spectra. However, we find a median value of $fc<1$, which suggests
that the LAMOST spectra have systematically redder spectral slopes
than the SDSS spectra for stars of the same spectral subtypes. We
suspect this is due to minor flux calibration differences between the
LAMOST and SDSS data reduction pipeline. The tails in the
distribution, on the other hand, are mainly due to LAMOST spectra with
instrumental issues or data reduction/calibration shortcomings. The
left wing (low $fc$ values) comprises stars in which there is an
over-subtraction of the sky lines, which makes the stars appear
systematically redder. The right wing of the distribution (high $fc$
values) comprises spectra which suffer from a known issue with the
blue channel calibration, making the stars appear bluer.

Interstellar reddening might also be affecting the distribution to
some extent, although the effect should be small given that most of
these M dwarfs are relatively nearby. To examine this possibility, we
separately plot in Figure~\ref{fc} the distribution of $fc$ values for
the relatively nearby (d$<$200pc) and the more distant (d$>$200pc) M
dwarfs in our list (see below). We find no significant offset between
the two distribution, which suggests that interstellar reddening is
negligible within our larger catalog, and that most of the $fc$ values
correct for instrumental errors in the spectrophotometry.

\subsection{Infrared color distribution}

One potentially serious source of contamination in any red dwarf
survey are background red giant stars, which also display molecular
bands of TiO, CaH, and VO, similar to the M dwarfs. To identify
possible red giants in our sample, we examined the 2MASS J-H and
H-K$_{\rm s}$ colors, of all the stars identified as M
dwarfs/subdwarfs by our pipeline. M giants occupy a distinct locus in
the (J-H, H-K$_{\rm s}$) diagram \citep{Bessell.1988,Lepine.2005b};
this is due to the development of water bands in M dwarfs which
depresses the flux in the H and K$_{\rm s}$ bands and make the dwarfs
appear bluer. In particular, M dwarfs always show $J-H<0.7$, while
giants usually have much redder $J-H$ colors.

The (J-H, H-K$_{\rm s}$) distribution of the stars identified as M
dwarfs/subdwarfs by the LAMOST pipeline is plotted in
Figure~\ref{jh_hk}. Stars are plotted in two separate subsets based on
their spectroscopic distance (see \S5 below). The nearest stars
($d<200pc$) occupy a relatively concentrated area which is remarkably
similar to the unreddened nearby M dwarfs locus in the (J-H, H-K$_{\rm
  s}$) diagram by \citet{Lepine.2005b}. The agreement suggest that
there is probably no significant interstellar reddening, no
significant contamination from red giant stars. The distribution of
the more distant stars ($d>200pc$), on the other hand, in the (J-H,
H-K$_{\rm s}$) diagram shows significantly more scatter, and also
displays a systematic average shift to redder infrared magnitudes,
with many stars having colors $J-H>0.7$. While this could be
indicating the presence of some red giant contaminants, this
possibility is not supported by the distribution of spectral indices.
After careful examination of all the LAMOST spectra for stars
with $J-H>0.7$, we conclude that the vast majority of the stars are
indeed M dwarfs and not M giants, with the notable exception of the
star shown in Figure~\ref{exp} and discussed above.

Instead, we suggest that the systematically redder colors may indicate
true interstellar reddening at least in some stars. This is supported
by the fact that the offset affects mostly the more distant stars in
our list. This reddening would have been corrected in our flux
renormalization procedure, and go unnoticed. In addition, the
increased scatter may be explained by the fainter infrared magnitudes
of the more distant M dwarfs. Overall, this indicates that giant star
contamination is not a significant problem in our M dwarfs/subdwarfs
catalog.

\section{Distance and Kinematics analysis}

\subsection{Spectroscopic distances}

Spectroscopic distances for the stars were estimated based on the
absolute infrared magnitude (MJ) to the spectral subtype (SpTy)
relationships. These relationships were re-calibrated using  a revised
census of nearby M dwarfs with both spectral types and  parallax
distances, which we assembled. We first identified M dwarf and M
subdwarf stars with parallax measurements documented in
\citet{Monet.1992,Harrington.1993,Altena.1995,Henry.2006,Leeuwen.2007,
Gatewood.2009,Lepine.2009,Smart.2010,Khrutskaya.2010,Riedel.2011,Jao.2011,Dittmann.2014},
with a smaller number of additional stars from L{\'e}pine et al.(2015,
in preparation). Within this sample, we identified 1459 stars with
known spectral subtypes as determined in
\citet{Reid.1995,Lepine.2003a,Lepine.2007,West.2011,Lepine.2013}; with
again some additional subtypes determinations from L{\'e}pine et
al.(2015, in preparation). The final calibration sample includes 1374
dM, 47 sdM, 21 esdM, and 17 usdM. The distribution of absolute J
magnitudes as a function of the spectral subtype is shown in
Figure~\ref{mag_spty}, with one panel for each of the metallicity
classes. For every star, random values of $\pm$ 0.1 subtype was added,
to spread out the data for more clarity.

The absolute magnitude to spectral subtype relationship for esdM and
usdM stars was determined with a simple fit of a linear function. For
sdM stars, we found a simple linear fit to be unsatisfactory due to an
apparent inflection point at later subtypes, and instead determined a
calibration using two linear segments. For M dwarfs, even the
two-segment fit proved unsatisfactory, and a three segment calibration
was determined instead. Those relationships are plotted in
Figure~\ref{mag_spty}. The corresponding values for the absolute J
magnitude as a function of subtype are listed in
Table~\ref{table3}. We estimate that this calibration provides values
of MJ with an uncertainty of about 0.7 magnitudes, yielding
spectroscopic distances with an accuracy of about 40\%.

Figure~\ref{dis_his} shows the distribution of spectroscopic distances
for stars in various ranges of spectral subtypes. Early-type red
dwarfs (K7-M2) are clearly detected over a much larger distance range
than mid-type objects (M3-M5), which is as one should expect from the
magnitude limit of the LAMOST survey. Only in the first distance bin
($d<100pc$) are mid-type M dwarfs significantly more numerous that
early-type ones. This suggests that the vast majority of M dwarfs that
will be observed and classified in the LAMOST survey will be
early-type M dwarfs at least to within about 500pc of the Sun. For
mid-type M dwarfs, the survey will only probe the volume within about
100pc of the Sun.

\subsection{Proper motions}

We obtained proper motions for most of the K/M dwarfs by searching the
PPMXL catalog of \citet{Roeser.2010} and the latest version of the
SUPERBLINK proper motion catalog \citep{Lepine.2005b,Lepine.2011}. We
used the positions of the stars in the LAMOST input catalog to search
for the positions in PPMXL of the LAMOST K/M dwarfs identified in our
pipeline, finding matches for 2,458 stars. Then we used the SUPERBLINK
catalog to match the rest of the 154 stars, which provided proper
motions for an additional 30 stars. Proper motions could not be
assigned for the remaining 124 sources. The proper motion range for
K/M dwarfs in our catalog is 1 mas yr$^{-1}$ $ < \mu < $ 416 mas
yr$^{-1}$, while the median value of the proper motion is 21 mas
yr$^{-1}$.

These proper motions can be used to validate our spectroscopic
distances, to the extent that the calculated transverse motion should
be consistent with the known, local kinematics of the Galactic disk.
The distribution of transverse velocities is plotted in
Figure~\ref{vt_dist} as a function of the spectroscopic distance. We
find that most of the stars have transverse motions in the 10-100 km
s$^{-1}$ range, which is indeed largely consistent with the local disk
kinematics. The more distant stars (according to our spectroscopic
distance measurements) are also found to have significantly smaller
proper motions, which is also as expected. We also do not see any
significant number of objects with unusually small transverse motions,
which is what one would expect if background M giants were
contaminating the sample, because misclassified giants would also have
significantly underestimated spectroscopic distances.

We measure the mean value of the transverse motion for stars in three
bins of spectroscopic distances ($d<400pc$, $400pc<d<800pc$,
$800pc<d<1200pc$), along with the dispersion about the mean. These are
plotted in Figure~\ref{vt_dist}. All values are consistent with a mean
velocity dispersion of $\sim40$km s$^{-1}$, but there is a hint of a
monotonic increase of the mean transverse velocity with distance. This
can be explained by uncertainties in the spectroscopic distances:
stars with underestimated distances will also have underestimated
transverse motion, and stars with overestimated distances will also
have overestimated transverse motions. Stars in the most distant bin
have mean transverse motions about $40\%$ larger than the stars in the
central bin (55 km s$^{-1}$ compared with 40 km s$^{-1}$); this
suggests that spectroscopic distances may be over- or under-estimated
by about $\pm$40\% as well. This caveat should be considered when
conducting a kinematical analysis of the M dwarfs detected in LAMOST.

\subsection{Distribution of space motions}

We could not obtain radial velocities for the K/M dwarfs we identified
due to unreliable wavelength calibration of the LAMOST commissioning
spectra. With the improvement of instrument and calibration method
in regular survey, the radial velocity estimates for K/M dwarfs
are expected to be available for upcoming data releases.

While the unavailability of radial velocity measurements limits the
amount of kinematical data, it is still possible to carry out basic
kinematics analysis from spectroscopic distances and proper motions
alone. This can be done using the statistical method described in
\citet{Lepine.2013}, in which it is assumed that the radial velocity
$R_{V} = 0$ for all K/M dwarfs, but different subsets are extracted to
study the velocity distribution in various projections of the local
(U,V,W) velocity space. Here we assume U is the component on motion
pointing in the direction of the Galactic center, V the component in
the direction of Galactic rotation, and W in the direction of the
north Galactic pole. The statistical method uses the fact that in some
specific directions on the sky, it is possible to calculate 2 of the 3
components of motion using distances and proper motions alone. For
example one can measure the components U and V from proper motions and
distances if the stars are in close proximity to the north or south
Galactic poles (because the radial velocity contributes almost
exclusively to the component W). Likewise the components of U and W
can be measured from distances and proper motions alone for stars located in
close proximity to the Galactic center and anti-center, and the
components of V and W can be measured from distances and proper
motions alone for stars located in close proximity to the apex and
antapex of the Galactic rotation ($l=90, b=0$; and $l=270, b=0$). The
approximation however breaks down for stars that are far from any of
those six points on the sky, but can be efficiently used in large
surveys for specific subsets.

Here we adopt the following approach: for every star in the sample, we
determine which of the six canonical points (as described above) is
closest to the star in question, and calculate the two components of
motion which can be approximated with distance and proper motion alone
for stars close to that point on the sky. Effectively this means that a
single pair of components of motion (either UV, UW, or VW) is
calculated for each star, is every case assuming that the radial
velocity is zero. Because the LAMOST commissioning data only covers
fields of low galactic latitudes, this means that all stars get either
estimates of (U,W) or of (V,W).

These distributions are shown in Figure~\ref{uvw}. It is interesting
that the (V,W) distribution clearly displays the expected local
asymmetric drift, with most stars having $V<0$km s$^{-1}$. The (U,W)
distribution, one the other hand, suggestively shows a non-isotropic
structure which could possibly be the signature of local streams,
although further analysis would be required, ideally with the
inclusion of radial velocity measurements, to confirm whether this
trend is real.

While this method does not provide complete (U,V,W) component for any
one of the stars, we still have sufficient numbers of stars
contributing information to each of the three velocity
components. Simply, stars with blanked values in one component of
motion are not used to calculate the statistical moments of that
component. The mean velocity and standard deviation for the (U,V,W)
components are therefore calculated to be:
\begin{displaymath}
 <U> =  -9.8 \pm 1.4~{\rm km s^{-1}}, \sigma_U = 35.6 \pm 6.1~{\rm km s^{-1}},
\end{displaymath}
\begin{displaymath}
 <V> = -22.8 \pm 0.7~{\rm km s^{-1}}, \sigma_V = 30.6 \pm 2.0~{\rm km s^{-1}},
\end{displaymath}
\begin{displaymath}
 <W> = -7.9 \pm 0.5~{\rm km s^{-1}}, \sigma_W = 28.4 \pm 3.3~{\rm km s^{-1}}.
\end{displaymath}
Where the uncertainties are estimated by bootstrap method.

Comparing with other reported results, such as \citet{hawley.1996} for
the PMSU survey, \citet{fuchs.2009} for the SDSS survey, and
\citet{Lepine.2013} for the SUPERBLINK survey, our (U,V,W) velocity
components are general consistent with those values. The agreement
further suggests that our spectroscopic distance estimates do not
suffer from systematic error, and are reliable enough to conduct
kinematic studies. This opens the prospect of using large numbers of
spectroscopically confirmed M dwarfs from LAMOST to perform massive
kinematics studies of the nearest 1kpc.

\section {Description of the catalog}

The complete catalog of 2,612 M dwarfs/subdwarfs is provided in
Table~\ref{table4}. Columns 1 and 2 catalog list celestial
coordinates at the 2000.0 epoch. Columns 3 to 6 list proper motions
along the Right Ascension and Declination, in milliarcseconds per
year, along with the measurement errors, when available. A flag
indicating the source of the proper motion is included in column
7. Columns 8 to 13 tabulate the infrared J, H, K$_{\rm s}$ magnitudes
of the counterparts in the 2MASS survey; 22 M dwarfs do not have
2MASS magnitude because they are too faint to be observed by
2MASS. Spectral subtypes which determined by our template-fitting
classification code are tabulated in column 14. Based on the spectral
subtypes, the estimated spectroscopic distances is listed in column 15.

\section{Conclusions}

We have successfully tested a template-fitting method to automatically
identify and classify the late-type K and M dwarfs in spectra from the
LAMOST survey. As an alternative to the classification software
'Hammer', used to analyze Sloan Digital Sky Survey spectra, our
procedure can perform more reliable spectral classification without
the need for time-consuming, visual inspection. Instead of relying on
the spectral indices measurement, we have assembled a set of
classification templates by combining spectra of M dwarfs/subdwarfs in
the Sloan Digital Sky Survey. The well-defined templates define a
temperature and metallicity grid, which reliably classifies the LAMOST
stars into subtypes (temperature) and metallicity subclasses by the
template-fit method. The method relies on a spectrophotometric
renormalization, which makes the fit less dependent of the observed
spectral energy distribution of the star, and more dependent on the
absolute and relative depths of the TiO, CaH, and other molecular
bands.

To improve on the classification of M dwarfs/subdwarfs, we introduce
subdivisions in the "metallicity" class system, going from 4 main
classes (dM, sdM, esdM, usdM) to 12 "metallicity" subclasses, with the
induction of three subdivisions for every metallicity class, labeled
``r'' for ``richer'', ``p'' for ``poorer'', and ``s'' for
``standard''. These subdivisions are included are suffixes to the
existing metallicity classes, thus providing a metallicity sequence
which runs [dMr, dMs, dMp, sdMr, sdMs, sdMp, esdMr, esdMs, esdMp,
usdMr, usdMs, usdMp], from the presumably most metal-rich to the more
metal-poor star. The term "metallicity" is used here in a suggestive
manner, and should not be understood in the strict sense. The
"metallicity" axis in our classification system is mainly dependent
on the TiO to CaH molecular band ratio, which is only assumed to be
correlated with the star's metallicity. Whether these "metallicity
classes" can be used to determine actual [Fe/H] values for each star
would first require a proper calibration and validation of the
classification grid, which is beyond the scope of this paper. In other
words, the proposed classification system should be used simply for
classification purposes at this time.

Using the LAMOST commissioning data acquired in 2009-2010, we
identified 2,612 M dwarfs/subdwarfs by our classification pipeline,
including 1,603 M dwarfs (dM), 52 metal-poor M subdwarfs (sdM), 5 very
metal-poor extreme subdwarfs (esdM), and 1 probable ultra metal-pool
subdwarf (usdM). Our quality controls and close examination of
individual stars indicate that the typical accuracy for our
template-fit method spectral classification to $\pm0.5$ subtypes and
to within 1 metallicity subclass (in the 12-subclass system).
A complete list of the 2,612 M dwarfs/subdwarfs is provided in a
table.

Contamination by background giant stars is shown to be negligible in
this subset. The distribution of infrared colors and the relative
strength of TiO and CaH molecular bands are inconsistent with
significant red giant contamination, although one star is indeed
identified (and flagged) as a probable giant.

To demonstrate proof-of-concept for using the LAMOST data for
kinematics studies, we estimated the spectroscopic distances and
heliocentric (U,V,W) velocity components to perform the kinematic
analysis for all M dwarfs/subdwarfs we identified. A spectroscopic
distance calibration is provided, which is based on collated parallax
data for M dwarfs and M subdwarfs in the literature. Transverse
motions are then calculated using proper motion data collected mainly
from the PPMXL and SUPERBLINK proper motion catalogs.

This preliminary work therefore demonstrates that future LAMOST survey
programs hold the potential to identify and classify very large
numbers of M dwarfs in all parts of the sky, and the data collected
will be of sufficient quality to obtain metallicity and kinematics
information on local M dwarfs to a distance of at least 500
parsecs. The anticipated data hold the promise to considerably expand
the statistics of the local M dwarf/subdwarf populations.

\acknowledgments

{\bf Acknowledgments}

This research was supported by `973 Program' 2014 CB845702, the Strategic
Priority Research Program "The Emergence of Cosmological Structures"
of the Chinese Academy of Sciences, Grant No. XDB09000000,  and
the National Science Foundation of China (NSFC)
under grants 11061120454 (PI:Deng), 11173044 (PI:Hou) and 11078006 (PI:Liu), by
the Shanghai Natural Science Foundation 14ZR1446900 (PI:Zhong),
by the Key Project 10833005 (PI:Hou), and by the Group Innovation Project
NO.11121062. This research was also supported by the United
States National Science Foundation under grants AST-0937523 (PI:Newberg),
AST-0607757 (PI:L{\'e}pine) and AST-0908406 (PI:L{\'e}pine).

Guoshoujing Telescope (the Large Sky Area Multi-Object Fiber Spectroscopic
Telescope LAMOST) is a National Major Scientific Project built by the Chinese
 Academy of Sciences. Funding for the project has been provided by the National
 Development and Reform Commission. LAMOST is operated and managed by the
  National Astronomical Observatories, Chinese Academy of Sciences.


\clearpage

\begin{deluxetable}{cccccccccccccrcccc}
\scriptsize

\tablecolumns{15}
\tablewidth{0pt}
\tablecaption{M dwarfs catalog,including astrometry,photometry,
	spectroscopic distances and estimated subtypes\tablenotemark{1}
\label{table4}}
\tablehead{
\colhead{RAJ2000\tablenotemark{2}} &
\colhead{DEJ2000} &
\colhead{pmRA\tablenotemark{3}} &
\colhead{pmDE} &
\colhead{e$\_$pmRA\tablenotemark{4}} &
\colhead{e$\_$pmDE} &
\colhead{fl\tablenotemark{5}} &
\colhead{J\tablenotemark{6}}&
\colhead{e$\_$J\tablenotemark{7}}&
\colhead{H} &
\colhead{e$\_$H} &
\colhead{K$_{\rm s}$} &
\colhead{e$\_$K$_{\rm s}$} &
\colhead{subtype\tablenotemark{8}} &
\colhead{Dist\tablenotemark{9}}  \\
\colhead{deg} &
\colhead{deg} &
\colhead{mas/yr} &
\colhead{mas/yr} &
\colhead{mas/yr} &
\colhead{mas/yr} &
\colhead{} &
\colhead{mag} &
\colhead{mag} &
\colhead{mag} &
\colhead{mag} &
\colhead{mag} &
\colhead{mag} &
\colhead{} &
\colhead{parsec}
}
\startdata
  18.561404 & 43.167682 &   7.4 & -13.9 &   3.9 &   3.9 & P &  14.23 &   0.03 &  13.56 &   0.03 &  13.37  &  0.04&   dKp7.5   &   485$\pm$145  \\
  17.613458 & 43.683223 &  -0.2 &  -7.9 &   4.0 &   4.0 & P &  15.67 &   0.07 &  15.00 &   0.08 &  14.67  &  0.09&   dMs1.5   &   651$\pm$195  \\
  16.418509 & 43.924841 &  -3.6 &  -2.9 &   4.1 &   4.1 & P &  15.79 &   0.07 &  15.02 &   0.06 &  14.78  &  0.08&  sdMs0.0   &   626$\pm$188  \\
   9.583018 & 40.944141 &  14.5 & -16.0 &   3.8 &   3.8 & P &  12.16 &   0.02 &  11.47 &   0.03 &  11.30  &  0.02&  sdMs0.0   &   117$\pm$ 35  \\
  16.169814 & 43.944604 &  -3.3 & -14.4 &   5.7 &   5.7 & P &  15.96 &   0.09 &  15.41 &   0.09 &  15.21  &  0.11&   dMs0.0   &   982$\pm$294  \\
  17.184962 & 43.590305 &   5.6 & -10.8 &   4.0 &   4.0 & P &  15.43 &   0.05 &  14.76 &   0.05 &  14.50  &  0.06&   dMr0.5   &   701$\pm$210  \\
   8.997519 & 41.646745 & -45.5 & -15.5 &   4.0 &   4.0 & P &  12.56 &   0.02 &  11.95 &   0.02 &  11.73  &  0.02&   dMs1.5   &   155$\pm$ 46  \\
  11.176808 & 41.781084 & -79.4 & -16.2 &   4.0 &   4.0 & P &  12.22 &   0.02 &  11.66 &   0.02 &  11.44  &  0.03&  sdMr1.0   &    85$\pm$ 25  \\
  11.422563 & 41.869348 & -27.0 & -23.2 &   4.0 &   4.0 & P &  12.98 &   0.02 &  12.36 &   0.02 &  12.19  &  0.02& esdKp7.0   &   130$\pm$ 39  \\
 123.894386 & 56.428029 &   1.5 &   8.8 &   4.0 &   4.0 & P &  12.78 &   0.02 &  12.08 &   0.02 &  11.97  &  0.02&   dKs7.5   &   248$\pm$ 74  \\
 124.445228 & 56.492803 &  12.5 & -98.0 &   4.0 &   4.0 & P &  11.88 &   0.02 &  11.28 &   0.02 &  11.13  &  0.02&   dKr7.0   &   180$\pm$ 54  \\
  18.985212 & 44.650124 & -24.4 &  -6.9 &   3.9 &   3.9 & P &  13.12 &   0.02 &  12.42 &   0.02 &  12.21  &  0.02&   dMs0.5   &   242$\pm$ 72  \\
  17.300816 & 47.198761 & -15.9 &  -6.3 &   4.0 &   4.0 & P &  14.41 &   0.04 &  13.76 &   0.04 &  13.64  &  0.04&   dKr7.0   &   579$\pm$173  \\
  16.860019 & 47.571482 & -12.8 &  -7.1 &   4.0 &   4.0 & P &  13.43 &   0.02 &  12.74 &   0.02 &  12.51  &  0.02&   dKr7.5   &   335$\pm$100  \\
  20.048857 & 45.172276 &  21.0 & -58.0 &   8.0 &   8.0 & S &  12.72 &   0.02 &  12.03 &   0.03 &  11.82  &  0.02&   dMs1.0   &   183$\pm$ 55  \\
  18.700087 & 43.153988 & 191.0 &-198.0 &   8.0 &   8.0 & S &  13.45 &   0.03 &  12.95 &   0.03 &  12.73  &  0.03&   dMr0.5   &   281$\pm$ 84  \\
  91.337318 & 23.565868 &  -2.0 & -59.0 &   8.0 &   8.0 & S &  12.45 &   0.02 &  11.76 &   0.02 &  11.53  &  0.02&   dMr3.5   &    81$\pm$ 24  \\
  17.400468 & 43.307743 &   --- &   --- &   --- &   --- & T &  15.56 &   0.06 &  14.90 &   0.06 &  14.53  &  0.07&   dKp7.5   &   895$\pm$268  \\
  18.608758 & 46.855343 &   --- &   --- &   --- &   --- & T &  12.94 &   0.03 &  12.29 &   0.02 &  12.09  &  0.02&  sdMr0.5   &   142$\pm$ 42  \\
  11.541000 & 41.583000 &   --- &   --- &   --- &   --- & L &   ---  &   ---  &   ---  &   ---  &   ---   &  --- &   dMr0.0   &        ---
\enddata
\tablenotetext{1}{The full version of this table is available in the
  electronic version of the Astronomical Journal. Twenty
  lines of the table are printed here to show the general layout.}
\tablenotetext{2}{Celestial coordinates in decimal degree,epoch 2000.0.}
\tablenotetext{3}{Proper motion in RA*cos(DEJ2000).}

\tablenotetext{4}{Mean error in pmRA*cos(DEJ2000).}
\tablenotetext{5}{Flags has the meaning: \\
       $~~~~\#$ P = Row of the astrometrical parameters
		   		   are from the PPMXL catalog.\\
       $~~~~\#$ S = Row of the astrometrical parameters
                      are from the SUPERBLINK catalog. \\
       $~~~~\#$ T = Coordinates parameters are from 2MASS catalog\\
       $~~~~\#$ L = Coordinates parameters are from LAMOST input catalog.}
\tablenotetext{6}{Infrared J, H, and K$_{\rm s}$ magnitudes from the
  2MASS catalog \citep{Skrutskie.2006}.}
  \tablenotetext{7}{Mean error of J, H, and K$_{\rm s}$ magnitudes from the
  2MASS catalog \citep{Skrutskie.2006}.}
\tablenotetext{8}{Estimated spectral subtype based on the template spectral fit.}
\tablenotetext{9}{Spectroscopic distance base on the absolute magnitude(Mj).}

\end{deluxetable}
\clearpage


\end{document}